# Quantum State Control in Optical Lattices

Ivan H. Deutsch
*Center for Advanced Studies, Department of Physics and Astronomy,*
*University of New Mexico, Albuquerque, NM 87131*

Poul S. Jessen
*Optical Sciences Center, University of Arizona, Tucson, AZ 85721*

We study the means to prepare and coherently manipulate atomic wave packets in optical lattices, with particular emphasis on alkali atoms in the far-detuned limit. We derive a general, basis independent expression for the lattice operator, and show that its off-diagonal elements can be tailored to couple the vibrational manifolds of separate magnetic sublevels. Using these couplings one can evolve the state of a trapped atom in a quantum coherent fashion, and prepare pure quantum states by resolved-sideband Raman cooling. We explore the use of atoms bound in optical lattices to study quantum tunneling and the generation of macroscopic superposition states in a double-well potential. Far-off-resonance optical potentials lend themselves particularly well to reservoir engineering via well controlled fluctuations in the potential, making the atom/lattice system attractive for the study of decoherence and the connection between classical and quantum physics.

PACS number(s): 32.80.Pj, 32.80.Qk, 73.40.Gk, 03.65.-w

## I. Introduction

One of the great challenges of modern science is to develop tools to prepare, manipulate and measure the quantum mechanical state of a physical system. Examples of systems in which quantum control is sought or has been accomplished are found in a wide range of fields. In physical chemistry, laser pulses are designed to direct chemical reactions along a desired pathway [1]. In quantum optics, nonclassical states of a single mode of the electromagnetic field have been prepared [2] and accurately measured [3], and several groups now pursue quantum state engineering of a single mode of a high-Q cavity [4]. In atomic physics, it has proved possible to control electronic orbital motion, and produce both quasi-classical and highly non-classical Rydberg wavepackets [5]. In ion traps, the exceptionally long-lived vibrational and hyperfine coherences, which originally inspired work on atomic clocks, have proven equally valuable in work on quantum state manipulation. In a series of recent experiments, the group of D. Wineland have demonstrated state preparation [6], state control [7] and even quantum "logic" gates [8] using trapped ions. Proposals for quantum logic have been made also in cavity QED [9]. This work constitutes an important step towards building a "quantum computer", in which algorithms are implemented as unitary transformations on a many-body quantum system [10].

Of particular fundamental interest in the context of coherent control are macroscopic superposition states, or "Schrödinger cats". The concept of incoherent evolution in such systems stemming from interaction with the environment forms the cornerstone of our understanding of the connection between classical and quantum physics [11]. For many years a paradigm for quantum coherence has been the observation of tunneling in macroscopic and mesoscopic systems [12]. The delocalized states resulting from tunneling over macroscopic distances are extremely susceptible to decoherence due to interaction with the classical environment, and it remains a key challenge to design systems for which these deleterious effects are minimized. Equally, it is of great importance to perform controlled studies of the effect of the environment on macroscopic and mesoscopic quantum states, so as to improve our understanding of the limits that apply when we attempt to evolve them in a quantum coherent fashion.

In this article we explore a new and promising system in which to study quantum state preparation, coherent control and the decoherence of macroscopic superposition states - neutral atoms trapped in an "optical lattice". Optical lattices are periodic potentials formed by the AC Stark shift (light shift) seen by atoms when they interact with a set of interfering laser beams [13]. In a suitable lattice formed by near-resonance light, laser cooling will quickly accumulate atoms in the few lowest bound states (Wannier states) of the individual optical potential wells, as demonstrated by resonance fluorescence and pump-probe spectroscopy [14]. Atoms prepared in this fashion can be transferred afterwards to a lattice formed by light detuned far from any atomic resonances [15], where they can be tightly bound in a nearly dissipation-free potential. Far-off-resonance optical lattices have been applied to the study of quantum chaos [16], and to the study of Bloch oscillations and Wannier Stark ladders [17]. When used to trap atoms in the tight binding regime, far-off-resonance lattices offer a realistic prospect of preparing pure quantum states, either by state selection or resolved-sideband Raman cooling [18].

A wide range of properties characterizing an optical lattice potential can be adjusted through laser beam geometry, polarization, intensity and frequency, and through the addition of static electric and magnetic fields. The richness and flexibility inherent to the multi-level atom-lattice interaction permits us to explore Hamiltonian evolution beyond the standard coupling of a spin-1/2 system to a harmonic oscillator (the Jaynes-Cummings model), which has been studied extensively in cavity QED [2] and ion traps [19]. Finite dissipation occurs in the lattice due to spontaneous photon scattering, but can be suppressed to an arbitrary degree in the far-detuned limit, as long as sufficient laser power is available to provide the desired potential. One can then design operations which evolve the atomic wave packet in a coherent fashion. Furthermore, once the intrinsic incoherent processes have been suppressed, dissipation may be re-engineered into the system in the form of well characterized fluctuations of the lattice potentials, allowing for a detailed study of the decoherence process. In either context, an important advantage of the atom/lattice system is the relative simplicity of the underlying interactions, which permits both the coherent and dissipative aspects of the evolution to be incorporated in a complete, yet tractable ab initio quantum theory.

To explore the evolution of macroscopic quantum coherence in a noisy environment, we specifically consider the tunneling of atoms in optical double-well potentials. The closely related process of tunneling from bound states to the continuum has been observed in an accelerated far off resonance standing wave [20], and found to show a signature of non-exponential decay [21]. There is a wealth of literature on the role of dissipation in tunneling [22], stimulated by the seminal paper by Calderia and Leggett [23] in their work relating to tunneling phenomena in Josephson junctions [24]. Since then, their ideas have been applied to a variety of systems in physics and chemistry [25]. As an example of relevance to our work here, quantum tunneling of atoms plays an important role in the dielectric properties of alkali-halide crystals and the acoustic properties of amorphous solids [26]. In the following we show how double potential wells can be designed through the combination of light shift and static magnetic dipole interactions. In this system we can observe tunneling of the *entire atom* through the optical wavelength-sized barrier separating the potential minima, and use this coupling to prepare Schrödinger cat states. From an experimental perspective we note that our double well potential has a built-in polarization gradient, so that tunneling is accompanied by a precession of the atom's angular momentum. This provides a label for left/right positions in the double well, and allows real-time observation of coherent tunneling oscillations as an oscillation in the magnetization - something that is typically not possible in a condensed matter system. A variety of similarly delocalized states have been produced in atom interferometers [27], in ion traps [7], and as a result of velocity-selective coherent population trapping [28].

The remainder of this article is organized as follows. In Sec. II we discuss how to design the three basic ingredients of coherent control in the atom/lattice system: potentials, coherent evolution, and state preparation. We focus largely on the alkalis, which have become the standard choice in experimental work with laser cooled atoms. Sec. II.A presents a general formalism which can be used to derive atomic potentials in the far-off-resonance limit, cast in a form which gives a clear physical picture. Sec. II.B explores coherent evolution operators based on coupling between magnetic sublevels, and Sec. II.C discusses how this coupling can be used to implement resolved-sideband Raman cooling and state preparation. In Sec. III we discuss how these ingredients may be applied to the study of quantum tunneling in double potential wells, and the creation of Schrödinger cat states. Finally, Sec. IV summarizes our results.

## II. Coherent Control and State Preparation.

The use of lasers to coherently control the internal state of atoms is a well established technique, many of whose methods were borrowed from nuclear spin resonance: Rabi flopping, rotation of the Bloch vector via $\theta$-pulses, adiabatic rapid passage, etc. These same techniques are readily applied to the manipulation of trapped ions [29] and cold neutral atoms [30], where the center-of-mass motion is included in the overall quantum state. One technique, which has been successfully applied to quantum state engineering with trapped ions, is to use a pair of laser fields to induce Raman transitions between the vibrational manifolds associated with a pair of hyperfine ground states. A similar approach can in principle be used for atoms trapped in optical lattices, if Raman coupling is introduced through the addition of separate coupling fields [31]. Note however, that optical potentials are crucially different from ion traps in one respect: the trapping potential depends strongly on the atomic internal state. This complexity will prevent a straightforward transfer of ion trap "technology", but at the same time it serves as an example of the greater richness of the atom/lattice system, which we can exploit as we develop new tools for coherent control. In this chapter we explore how appropriate Raman coupling terms can be designed into the optical lattice potential itself, and subsequently be used as building blocks for coherent evolution operators. Our main goal here is to present the physical concepts involved in this design process, something which is most clearly done in the context of simple 1D and 2D lattice configurations. In these 1D and 2D lattices we will apply the terms "quantum state" and "coherent evolution" to the internal state and quantized motion in the lattice directions only; motion in the unbound direction(s) is separable, and can be ignored.

## A. Designing Atomic Potentials.

A variety of potentials for cold atoms can be designed via their interaction with the electromagnetic field, including optical fields (light shifts), static magnetic fields (Zeeman shifts) and static electric fields (Stark shifts). We here specialize to the case of low intensity monochromatic light, $\mathbf{E}_L(\mathbf{x},t) = \mathrm{Re}\left[\mathbf{E}_L(\mathbf{x})e^{-i\omega_L t}\right]$, and static magnetic fields, so that the potential for atoms in the ground state reads

$$\hat{U}(\mathbf{x}) = -\mathbf{E}_L^*(\mathbf{x}) \cdot \hat{\alpha} \cdot \mathbf{E}_L(\mathbf{x}) - \hat{\mu} \cdot \mathbf{B}, \tag{1}$$

where $\hat{\alpha} = -\sum_e \hat{\mathbf{d}}_{ge}\hat{\mathbf{d}}_{eg} / \hbar\Delta_{ge}$ is the atomic polarizability tensor operator (in the far-off resonance limit), with $\Delta_{ge}$ the detuning from the $|g\rangle \to |e\rangle$ resonance, and where $\hat{\mathbf{d}}_{eg}$ is the electric dipole operator between these states; $\hat{\mu} = \hbar\gamma\hat{\mathbf{F}}$ is the magnetic dipole operator, with $\gamma$ the gyromagnetic ratio and $\hat{\mathbf{F}}$ the total angular momentum operator.

To illustrate some of the features of the potential which can be easily controlled, consider an atom driven on a $|J=1/2\rangle \to |J'=3/2\rangle$ transition by a one dimensional optical lattice, produced by a pair of red-detuned, counterpropagating plane waves with amplitude $E_1$ and angle $\theta$ between their linear polarizations. We will refer to this configuration as "1D lin-angle-lin"; in the special case $\theta = \pi/2$ it reduces to the familiar 1D lin⊥lin lattice. Near resonance laser cooling in lattice configurations with arbitrary $\theta$ has been studied by Finkelstein, Berman and Guo [32] and by Taieb et al. [33]. The optical field can be written as a superposition of opposite helicity standing waves,

$$\mathbf{E}_L(z) = \sqrt{2}E_1\left\{-e^{-i\theta/2}\cos(k_L z + \theta/2)\mathbf{e}_+ + e^{i\theta/2}\cos(k_L z - \theta/2)\mathbf{e}_-\right\}, \tag{2}$$

for a convenient choice of relative phase between the beams. The potential, given by Eq. (1), is then

$$\hat{U}(z) = -\frac{2U_1}{3}\left\{2\big(1+\cos\theta\cos(2k_L z)\big)\hat{I} + (\sin\theta\sin(2k_L z))\hat{\sigma}_z\right\} - \frac{\hbar}{2}\gamma\mathbf{B}\cdot\hat{\sigma}, \tag{3}$$

where $U_1$ is the light shift produced by a single beam of amplitude $E_1$, driving a transition with unit Clebsch-Gordan coefficient (henceforth the "single beam" light shift). The operators $\left\{\hat{I}, \hat{\sigma}_i\right\}$ are the identity and Pauli spin operators in the ground state manifold. Note that a magnetic field along the laser axis (*z*-direction) does not break the rotational symmetry of the potential; transverse magnetic fields break this symmetry and thus establish coherences between the magnetic sublevels.

Adiabatic potentials can now be found by diagonalizing $\hat{U}(z)$. Fig. 1 shows the adiabatic bi-potentials associated with different lattice configurations. In the absence of a transverse **B**-field, $\hat{U}(z)$ is diagonal in the eigenstates of $J_Z$. Varying $\theta$ changes the peak-peak modulation depth $U_p$ of the potential, and the distance $\Delta z$ between the $|m=+1/2\rangle$ and $|m=-1/2\rangle$ potential wells (Fig. 1b),

$$U_p = \frac{4}{3} U_1 \sqrt{3\cos^2\theta + 1}\,, \qquad k_L \Delta z = \tan^{-1}\left(\frac{\tan\theta}{2}\right), \tag{4}$$

while changing $B_Z$ shifts the minima of these wells (Fig. 1c). By adding a transverse magnetic field we break the degeneracy of the bi-potential at positions of linearly polarized light (Fig. 1d). Thus, by choosing the appropriate angle between the laser polarizations and an appropriate **B**-field, we can design a lattice of double well potentials with adjustable barrier and asymmetry. In Sec. III we will discuss the use of this potential to study quantum tunneling and macroscopic superposition states.

Though the $|J=1/2\rangle \to |J'=3/2\rangle$ system is useful for gaining physical intuition, the above results must be generalized to atoms with more complicated internal structure. Here we concentrate on the alkalis, and particularly Cs. For atoms optically pumped into a given hyperfine ground state, and having a multiplet of hyperfine excited states,

$$\hat{\alpha} = -\sum_{F'} \frac{P_F\,\hat{\mathbf{d}}\,P_{F'}\,\hat{\mathbf{d}}\,P_F}{\hbar\Delta_{F,F'}}, \tag{5}$$

where $P_F = \sum_m |F,m\rangle\langle F,m|$, $P_{F'} = \sum_{m'} |F',m'\rangle\langle F',m'|$ are projection operators onto the ground and excited hyperfine levels, respectively. As shown in Appendix (A), the components of the polarizability tensor in the spherical basis, $\hat{\alpha}_{q',q} = \mathbf{e}^*_{q'} \cdot \hat{\alpha} \cdot \mathbf{e}_q$, can be written as

$$\hat{\alpha}_{q',q} = \tilde{\alpha} \sum_{F'} \frac{\Delta_{F_{\max},F'_{\max}}}{\Delta_{F,F'}} f_{F'F} \sum_{m_F} c^{F',m+q}_{F,m+q-q'}\, c^{F',m+q}_{F,m} |F,m+q-q'\rangle\langle F,m|\,. \tag{6}$$

In this expression $f_{F'F}$ are the relative oscillator strengths for decay $|F'\rangle \to |F\rangle$, $\Delta_{F,F'}$ is the detuning of the laser from this resonance, $|F_{\max} = J+I\rangle$ and $|F'_{\max} = J'+I\rangle$ are the "stretched states", and $c^{F',m'}_{F,m}$ are the Clebsch-Gordan coefficients for the $|F,m\rangle \to |F',m'\rangle$ dipole transition. The characteristic polarizabilty scalar for the $|J\rangle \to |J'\rangle$ transition is defined as

$$\tilde{\alpha} \equiv \frac{\left|\langle J' \| d \| J \rangle\right|^2}{\hbar \Delta_{F_{\max}, F'_{\max}}} \ , \tag{7}$$

where $\langle J' \| d \| J \rangle$ is the dipole operator reduced matrix element. For this more complex system, an optical lattice with polarization gradients will generally establish coherences between the ground state magnetic sublevels via stimulated Raman transitions. These coherences (in conjunction with the externally imposed magnetic field) can be used to control the state of the atomic wave packet as discussed below.

Because we are interested in coherent evolution of the atomic state, the lattice should be detuned as far from resonance as possible. It is therefore important to examine the nature of the potential in the limit that the detuning is much larger than the hyperfine splittings. In that case Eq. (5) reduces to

$$\hat{\alpha} \approx P_F \, \hat{\alpha}(J \to J') \, P_F, \tag{8}$$

where $\hat{\alpha}(J \to J')$ is the polarizability tensor of the $|J\rangle \to |J'\rangle$ transition. Thus, for the alkalis, the very far off-resonance optical lattice has properties quite similar to the familiar $|J=1/2\rangle \to |J'=3/2\rangle$ transition. The operator $\hat{\alpha}(J \to J')$ is a rank 2 tensor which can be written as a sum of irreducible tensors of rank 0, 1, and 2. Because it acts on a two dimensional Hilbert space, in which any operator can be written as a superposition of scalar and vector operators $\left\{\hat{I}, \hat{\sigma}_i\right\}$, it follows that the irreducible rank 2 component must vanish exactly. In Appendix (B) we show that

$$\hat{\alpha}_{ij}(J \to J') = \tilde{\alpha} \left( \frac{2}{3} \delta_{ij} \, \hat{I} - \frac{i}{3} \varepsilon_{ijk} \, \hat{\sigma}_k \right). \tag{9}$$

Let us now express the lattice field as $\mathbf{E}_L(\mathbf{x}) = \mathrm{Re}(E \, \vec{\varepsilon}_L(\mathbf{x}) e^{-i\omega_L t})$, where $\vec{\varepsilon}_L(\mathbf{x})$ is the local polarization (not necessarily unit norm), and where we have factored out some conveniently chosen amplitude $E$. For concreteness, we will assume that the field is formed by a set of equal amplitude beams, in which case it is most convenient to factor out the single-beam amplitude $E_1$. The optical potential can then be written in the compact from

$$\hat{U}(\mathbf{x}) = U_J(\mathbf{x}) \, \hat{I} + \mathbf{B}_{eff}(\mathbf{x}) \cdot \hat{\sigma}$$
$$U_J = -\frac{2}{3} U_1 \left|\vec{\varepsilon}_L(\mathbf{x})\right|^2 \ , \quad \mathbf{B}_{eff}(\mathbf{x}) = \frac{i}{3} U_1 \left( \vec{\varepsilon}_L^{\,*}(\mathbf{x}) \times \vec{\varepsilon}_L(\mathbf{x}) \right), \tag{10}$$

where $U_1 = \tilde{\alpha} E_1^2 / 4$ is the single beam light shift. In other words, the light-shift potential is equivalent to a shift proportional to the local intensity of the field and independent of the hyperfine state of the atom, and an effective static magnetic field whose magnitude and direction depend on the local ellipticity of the laser polarization [34]. Using Landé's Projection Theorem, for the "stretched" ground hyperfine level with $F = I + J$, Eq. (10) yields

$$\hat{U}_F(\mathbf{x}) = U_J(\mathbf{x})\,\hat{I} + \mathbf{B}_{eff}(\mathbf{x}) \cdot \frac{\hat{\mathbf{F}}}{F}. \tag{11}$$

This expression is useful for interpreting the physical nature of the potential, and also very convenient for calculations because it is basis independent.

From Eqs. (10) and (11) we can make the following observations. In the limit of infinite detuning, coherences $|F,m\rangle \leftrightarrow |F,m\pm 2\rangle$ go to zero. If the light field is everywhere linearly polarized, the effective magnetic field vanishes, and the light shift is independent of the magentic substate of the atom, $\hat{U}_F(\mathbf{x}) = U_J(\mathbf{x})\,\hat{I}$. For fields with arbitrary ellipticity, polarization in the *x*-*y* plane gives rise to effective longitudinal **B** fields; the combination of $\pi$-polarized and $\sigma$-polarized light yields an effective transverse field. Unlike a true, externally applied, static magnetic field, the effective magnetic field can vary spatially with a period on the order of the optical wavelength. This dependence will be important in designing potentials that generate coherences between given atomic vibrational states.

**B. Designing Coherent Evolution Operators.**

Consider a typical 1D lin⊥lin optical lattice for Cs atoms, formed by light detuned $2000\,\Gamma$ to the red of the $6S_{1/2}(F=4) \rightarrow 6P_{3/2}(F'=5)$ resonance, with a single-beam light shift $U_1 = 150\,E_R$. The bandstructure is easily calculated [35], and is shown in Fig. 2. Without further perturbation tunneling between neighboring wells is negligible, i. e. the system is in the tight binding regime, and we can consider each lattice site as an independent potential well. This conclusion holds also for lattices of similar depth in higher dimensions. In this limit an appropriate description is given by the set of Wannier states which constitute an orthonormal basis within each well; Wannier states associated with different lattice sites are also orthogonal. Coherences between magnetic sublevels can arise due to Raman coupling terms in the lattice potential, Eq. (1), and the Wannier states in general become spinors [36]. Most optical lattices are designed to have pure helicity at the points of maximum light shift, in which case the most deeply bound states have negligible admixture of *m*-states. In that case the Wannier spinor is approximately a local harmonic oscillator state for the given diabatic potential, $|n,m\rangle \equiv |\Phi_n^{(m)}\rangle|F,m\rangle$. A notable exception occurs when a pair of states in the vibrational manifolds $\{|n,m\rangle\}$, $\{|n',m'\rangle\}$ are nearly degenerate, and coherent mixing via stimulated Raman transitions becomes resonantly enhanced, as shown by Courtois [37]. By applying a longitudinal magnetic field one can use the Zeeman shift to tune different levels into and out of Raman-resonance at will, and design various coherent evolution operators ($\pi$-pulses, adiabatic-rapid-passage, etc.) in a manner closely analogous to the methods applied in ion traps. Fig. 3 illustrates this procedure for Cs in a 1D lin⊥lin far-off-resonance lattice. In the following we examine the types of Raman coupling available in some representative lattice geometries.

In general the electromagnetic field can induce coherences of the form $|m'\rangle \leftrightarrow |m+\Delta m\rangle$, $\Delta m = \pm 1, 2$, through stimulated Raman transitions of the type $\pi \leftrightarrow \sigma_\pm$ and $\sigma_+ \leftrightarrow \sigma_-$. For alkali atoms at large but finite detuning both types of coupling occur, though in the infinite detuning limit only coherences $\Delta m = \pm 1$ persist as shown by Eq. (11). To establish the important scaling laws we first consider the $\Delta m = \pm 2$ coherences in a 1D lin⊥lin lattice. In a two-level picture, the strength of the Raman coupling is determined by the off-diagonal matrix element of the light shift operator,

$$U_R \approx U_1\, \beta_{2,4} \langle n' = 1, m = 2 | \sin(2k_L z) | n = 0,\ m = 4 \rangle. \tag{12}$$

The parameter $\beta_{2,4}$ determines the coupling of the internal degrees of freedom, whereas the matrix element determines the effective Franck-Condon overlap. To lowest order in the Lamb-Dicke parameter, $\eta = k_L z_0 = \sqrt{E_R/\hbar\omega_{osc}}$, where $z_0$ is the ground state variance, and ignoring the difference in curvature of the $m = 4, 2$ wells, the coupling matrix element is $U_R \approx 2U_1\, \beta_{2,4}\, \eta$. The full detuning dependence of the coupling constant can be obtained from Eq. (6),

$$\beta_{2,4} = \sum_{F'} f_{4F'} \left( \frac{\Delta_{4,5}}{\Delta_{4,F'}} \right) c_{4,2}^{F',3}\, c_{4,4}^{F',3} = \frac{\sqrt{7}}{360} \left( 16 - 21 \frac{\Delta_{4,5}}{\Delta_{4,5} + \delta_{5,4}} + 5 \frac{\Delta_{4,5}}{\Delta_{4,5} + \delta_{5,3}} \right), \tag{13}$$

where $\delta_{F_1',F_2'}$ is the splitting between excited state hyperfine levels $F_1'$ and $F_2'$. According to Landé's interval rule, the splitting between hyperfine levels $F'$ and $F'-1$ is proportional to $F'$, so $\delta_{5,4} \approx 5*\delta$ and $\delta_{5,3} \approx 9*\delta$, where $\delta \approx 10\Gamma$ for Cs. Expanding Eq. (13) to lowest order in $\Gamma / \Delta_{4,5}$, gives the asymptotic expression,

$$\beta_{2,4} \approx 4.4 \frac{\Gamma}{\Delta_{4,5}}. \tag{14}$$

Coherent manipulation of the atomic state requires that the timescale $\hbar/U_R$ for coherent evolution be shorter than the lifetime of the Raman coherence between states $|n', m = 2\rangle$ and $|n, m = 4\rangle$, which is dominated by the decay of the state $|n', m = 2\rangle$ due to optical pumping, and thus of order $\gamma_s^{-1}$ (the inverse photon scattering rate). We can then define a figure-of-merit for coherent manipulation,

$$\kappa \equiv \frac{U_R}{\hbar\gamma_s} \approx \eta\beta_{2,4}\frac{\Delta_{4,5}}{\Gamma} \approx 4.4\ \eta, \tag{15}$$

which should be much larger than unity. We see that for Raman coupling $\Delta m = \pm 2$, the figure-of-merit is of the same order as the Lamb-Dicke parameter, which is small by assumption. Thus these coherences will generally not be useful for full coherent control.

One is left with the possibility to induce coherences $\Delta m = \pm 1$ through the addition of an external transverse magnetic field or via stimulated $\pi \leftrightarrow \sigma_\pm$ Raman transitions. The coupling matrix element is then

$$U_{m,m\pm 1} = \frac{\langle\{n'\}, m\pm 1|(B_x^{tot}(\mathbf{x}) \mp iB_y^{tot}(\mathbf{x}))\hat{F}_\pm|\{n\},m\rangle}{2F}, \tag{16}$$

where $\mathbf{B}^{tot}$ is the sum of the external and the effective magnetic field given by Eq. (10). Clearly, strong coupling of the internal degrees of freedom is guaranteed, but the Franck-Condon overlap depends on the local spatial symmetry of the total "magnetic field". An external transverse magnetic field creates a spatially uniform coupling, which does not connect states of different parity, $|m,n\rangle$, $|m\pm 1, n\pm 1\rangle$, located at a given lattice site, but is useful in other contexts. In Sec. III we employ external transverse magnetic fields to couple states localized at different lattices sites of a 1D lin-angle-lin lattice, thereby inducing quantum tunneling. In contrast to an external field, the effective magnetic field provided by the lattice light field can be designed to provide both even and odd parity coupling. In the infinite detuning limit the off-diagonal element of the light shift operator, Eq. (11), becomes

$$U_{m,m\pm 1} = -\frac{U_1}{3\sqrt{2}}\frac{\sqrt{F(F+1)-m(m\pm 1)}}{F}\langle\{n'\}|\left(\varepsilon_\pi^*(\mathbf{x})\varepsilon_{\sigma_\pm}(\mathbf{x}) + \varepsilon_{\sigma_\mp}^*(\mathbf{x})\varepsilon_\pi(\mathbf{x})\right)|\{n\}\rangle \tag{17}$$

where $\varepsilon_{\sigma+}(\mathbf{x})$, $\varepsilon_{\sigma-}(\mathbf{x})$ and $\varepsilon_\pi(\mathbf{x})$ are the normalized electric field amplitudes of the lattice $\sigma_\pm$ and $\pi$-polarized components, and $\{n\}$ are the vibrational quantum numbers. From Eqs. (10) and (17) one can deduce a few general aspects of the lattice spatial symmetries. The Raman coupling terms will have nodes where the light is either purely $\sigma_\pm$ or $\pi$ polarized. Generally the representation of the light-shift operator in terms of diabatic potentials and off-diagonal Raman couplings depends on the choice of quantization axis. The exception is at positions where the light is linearly polarized. At these positions the effective magnetic field vanishes, irrespective of the quantization axis, and thus all the diabatic potentials are degenerate and the Raman coupling is zero. These features are illustrated in the examples below.

The 1D lin-angle-lin class of optical lattices include all configurations that can be formed by a set of counterpropagating plane waves. Thus the light fields of purely one-dimensional lattices can always be decomposed in $\sigma_\pm$ components, and it follows that one cannot introduce $\pi \leftrightarrow \sigma_\pm$ type Raman coupling. We can however misalign slightly the direction of propagation of one on the beams, so that a small component of polarization lies along the axis of the standing wave. For an arbitrary pair of cross polarized lasers (not necessarily counterpropagating), the optical potential in the very far-off-resonance limit is

$$\hat{U}(\mathbf{x}) = -\frac{4}{3}U_1\,\hat{I} - \frac{2}{3F}U_1 \sin\left[\left(\mathbf{k}_1 - \mathbf{k}_2\right)\cdot\mathbf{x}\right](\mathbf{e}_1\times\mathbf{e}_2)\cdot\hat{\mathbf{F}}. \tag{18}$$

Choosing the quantization axis along $\mathbf{k}_1 - \mathbf{k}_2$, Eq. (18) implies that both the diabatic and off-diagonal potentials have the same spatial dependence, and thus the off-diagonal coupling is always an even parity operator with respect to well-center.

We have much greater flexibility to design the coupling potential in higher dimensional lattices. Consider, for example, a 2D lattice formed by three coplanar laser beams of equal intensity and linearly polarized in that plane as shown in Fig. 4. Grynberg et al. used this geometery (with lasers near atomic resonance) to cool and trap Cs atoms in the Lamb-Dicke regime [38]. A slight rotation of the linear polarization of one beam out of the *x*-*y* plane introduces a π-component, and thus stimulated $\pi \leftrightarrow \sigma_\pm$ couplings. In contrast to the 1D geometry of Eq. (18), the 2D configuration permits us to vary the phase of the π-component independently from the phase of the σ-components, and thus allows us to optimize the desired couplings. Consider a π-polarized component with amplitude $E_\pi$ and a relative phase $\exp(i\varphi)$ to the lattice beam propagating in the $-y$ direction (in general this would correspond to some elliptical polarization of that beam). For a choice of relative phase between the beams which puts the maximum of the $\sigma_+$ polarized light at the origin, the lattice electric field is

$$\begin{aligned} E_L(\mathbf{x}) = \frac{E_1 e^{-iky}}{\sqrt{2}}\Big[&-\mathbf{e}_+\left\{1 + 2e^{iK_y y}\cos(K_x x)\right\} \\ &+ \mathbf{e}_-\left\{1 + 2e^{iK_y y}\cos(K_x x - 2\theta)\right\}\Big] + \mathbf{e}_\pi E_\pi e^{-iky} \end{aligned}, \tag{19}$$

where $K_x = k\sin\theta$, $K_y = k(1+\cos\theta)$. The effective field governing the coupling $|m = F\rangle \leftrightarrow |m = F-1\rangle$ is

$$\begin{aligned} B_x^{eff} + iB_y^{eff} = -\frac{2U_1}{3}\frac{E_\pi}{E_1}\Big\{&2\sin\theta\sin(K_x x - \theta)\cos(K_y y - \varphi) + \\ &+ 2i\cos\theta\cos(K_x x - \theta)\sin(K_y y - \varphi) - i\sin\varphi\Big\} \end{aligned}, \tag{20}$$

where $U_1$ is the single beam light shift. We can now use Eqs. (16) and (20) to calculate the matrix element $U_{m=F,m'=F-1}$ of the Raman coupling. Figs. 4b,c show cuts of the diabatic potentials and coupling matrix elements along the *x* and *y* directions for $\theta = \pi/3$, and reveal that the Raman coupling has both even and odd terms along both $x$ and $y$ , whose relative magnitude and phase can be controlled through the ellipticity of the beam polarization.

Expanding around the minimum of the potential well at the origin and making the harmonic approximation for the vibrational levels, we find to first order in the small parameters $kx$, $ky$,

$$U_{F,F-1} \approx -\frac{U_1}{\sqrt{2F}}\frac{E_\pi}{E_1}\left\langle \left\{n_x', n_y'\right\}, F-1 \left\| \left[ -e^{i\varphi} + \tfrac{1}{2} e^{-i\varphi} kx + i\left(e^{i\varphi} - \tfrac{1}{2} e^{-i\varphi}\right) ky \right] \right\| \left\{n_x, n_y\right\}, F \right\rangle . \tag{21}$$

Maximal coupling of the odd parity states results for $\varphi = \pi/2$, in which case the coupling matrix elements for vibrational change of one quantum along $x$ and $y$ are

$$U_{F,F-1}^{(x)} \approx iU_R\sqrt{n_x} \ , \quad U_{F,F-1}^{(y)} \approx 3U_R\sqrt{n_y} \ , \quad U_R = \frac{U_1}{2\sqrt{2F}}\frac{E_\pi}{E_1}\eta \tag{22}$$

where the Lamb-Dicke Parameter is

$$\eta = \sqrt{\frac{E_R}{\hbar\omega_{osc}}} = \left(\frac{2}{15}\frac{E_R}{U_1}\right)^{1/4} . \tag{23}$$

Computing the figure-of-merit for coherent manipulation we now find

$$\kappa \equiv \frac{U_R}{\hbar\gamma_s} \approx \frac{0.047}{\sqrt{F}}\frac{E_\pi}{E_1}\frac{|\Delta|}{\Gamma}\left(\frac{E_R}{U_1}\right)^{1/4} . \tag{24}$$

If we consider for example Cs ($F = 4$) in a lattice with $U_1 = 25\,E_R$, $\Delta = -10^4\,\Gamma$ and $E_\pi = 0.5\,E_1$, then we obtain $\kappa \approx 53$. Even more favorable figures-of-merit can be obtained at larger detunings, provided that sufficient laser power is available.

## C. State Preparation.

We are generally interested in the coherent evolution of quantum systems initially prepared in a pure state. The term "pure state" is used here to describe an ensemble of identically prepared atoms (of order $10^6$) localized at *different* lattice sites; in this case no single quantum state is macroscopically occupied. The preparation of a pure state can be accomplished by state selection, by dissipative cooling of the system to its ground state, or by a combination thereof. State selection techniques demonstrated in optical lattices include gravitational [39] and inertial [20] acceleration in shallow lattices supporting only one bound state, and in work with metastable noble gas atoms, selective quenching of vibrationally excited states [40]. These methods are most useful if a substantial fraction of the atoms initially occupy the desired quantum state. In a 1D lin⊥lin lattice this situation is readily achieved by near-resonance Sisyphus cooling. In the case of Cs atoms it has been found that a longitudinal magnetic field allows the preparation of up to 28% of the total population in the vibrational ground state associated with a single stretched state [41]. The addition of a transverse magnetic field serves to enhance this population due to the induced coherences between the magnetic sublevels; numerical simulations have shown that the proper combination of transverse and longitudinal magnetic fields should allow the population to be increased to 45%. As demonstrated in [15], atoms prepared in this fashion can be transferred to a far-off-resonance lattice, with close to unit efficiency and no significant increase in vibrational excitation.

In contrast to the one-dimensional case, the vibrational degeneracy occurring in two and especially three dimensions prevents one from obtaining useful ground state populations solely with Sisyphus cooling in a near-resonance lattice. In that case, preparation of a pure state will require additional cooling after the atoms have been transferred into a far-off-resonance lattice. Because we are primarily interested in the preparation of well localized Wannier states, the most efficient method is resolved-sideband cooling. In the Lamb-Dicke regime this technique in principle allows for the removal of one quantum of vibrational energy every few oscillation periods. Thus the rate of vibrational excitation must be well below the frequency of oscillation, $d\bar{n}/dt << \omega_{osc}$. In an optical lattice, heating is dominated by photon scattering; in the harmonic approximation the condition becomes $\omega_{osc} >> \eta^2 \gamma_s$, or equivalently $\left(\hbar\omega_{osc}/E_R\right)^2 >> \hbar\gamma_s/E_R$. This requirement is easily met by several of orders of magnitude, even in lattices detuned by only a few thousand linewidths [15].

In the following we explore a scheme for resolved-sideband Raman cooling which is based on transitions from states $|n, m = F\rangle$ in the vibrational manifold of the stretched state, to states $|n-1, m = F - \Delta m\rangle$ in the vibrational manifold of another magnetic sublevel, as illustrated in Fig. 5. As discussed in Sec. II.B, in a 1D lin-angle-lin lattice, odd-parity coupling operators are available only with the $\Delta m = 2$ type transitions, while 2D and 3D geometries allow $\Delta m = 1, 2$ depending on the details of the lattice geometry. Relaxation back to the states $|n-1, m = F\rangle$ is provided by optical pumping, resulting in a net loss of nearly one quantum of vibrational excitation per cooling cycle. It is important to note here that the required Raman coupling strength is much less for sideband cooling than for fully coherent population transfer, because the process which destroys Raman coherence, i. e. optical pumping $|n, m = F - \Delta m\rangle \rightarrow |n, m = F\rangle$, is also the process that accomplishes sideband cooling. In that case it is only necessary that the timescale for population transfer, $\hbar/|U_R|$, be much shorter than the timescale for vibrational excitation, i. e.,

$$\frac{|U_R|}{\hbar\, d\bar{n}/dt} = \kappa' >> 1. \tag{25}$$

To leading order in $\eta$, the rate of vibrational excitation is $d\bar{n}/dt = \gamma_s\left(\Delta k z_0\right)^2$, where $(\hbar\Delta k)^2$ is the mean-squared momentum transfer in a photon scattering event, obtained by averaging the momentum components along the lattice directions over the dipole emission pattern. Using Eqs. (12) and (14) for the Raman coupling strength, we can then compute the sideband cooling figure of merit $\kappa'$ in different types of lattices. First we consider a simple 1D lin⊥lin lattice with $\Delta m = 2$, for which we find

$$\kappa' \approx 9.1\left(\frac{U_1}{E_R}\right)^{1/4}. \tag{26}$$

For a large, but still realistic depth $U_1 = 500\ E_R$ we obtain $\kappa' \approx 43$. The figure-of-merit is further improved in the 2D lattice configuration of Fig. 4. As discussed in Sec. II.B, the polarization of one lattice beam is made elliptical in order to generate a strong Raman coupling, as given by the matrix element, Eq. (22). This yields the following figures of merit for the *x* and *y* directions,

$$\kappa'_x \approx 0.17\,\frac{E_\pi}{E_1}\frac{|\Delta|}{\Gamma}\left(\frac{U_1}{E_R}\right)^{1/4}, \qquad \kappa'_y = 3\ \kappa'_x. \tag{27}$$

For realistic parameters $U_1 = 45\ E_R$, $\Delta = -4000\Gamma$ and $E_\pi/E_1 = 0.5$ we obtain $\kappa'_x \approx 880$ and $\kappa'_y \approx 2.6\times10^3$.

To explore the prospects for resolved-sideband Raman cooling in more detail, we consider a simplified numerical model for cooling of Cs atoms in a 1D lin⊥lin far-off-resonance lattice. We restrict the system to the two vibrational manifolds $\{|n, m=4\rangle\}$ and $\{|n', m=2\rangle\}$, approximated by harmonic oscillators with eigenfrequencies $\hbar\omega_4 = 4\sqrt{U_1E_R/3}$ and $\hbar\omega_2 = 2\sqrt{2U_1E_R/3}$ (Fig. 5b). To accomplish one step of the cooling, a longitudinal magnetic field $B_z$ is used to shift a particular pair of states $n - n' = 1$ into degeneracy; in the terminology of sideband cooling this corresponds to tuning the Raman coupling to the red sideband. Rabi oscillations between a pair of states are always overdamped, $U_R/\hbar\gamma_s \le 1$, and thus the largest rate of population transfer is achieved by continuous resonant excitation. Because the coupled potential wells have different oscillation frequencies, resonant excitation on the red Raman sideband can be achieved only for one pair of levels, $|n-1, m=2\rangle$, $|n, m=4\rangle$, at a time. We ignore any accidental degeneracies which may occur simultaneously between higher lying states, since these are not significantly populated for the vibrational temperatures of interest.

As the Raman coupling is small compared to the oscillation frequency (resolved-sideband limit), it can be treated as a perturbation and we can separate our system into a series of $n$ two-level systems $\{|n-1, m=2\rangle, |n, m=4\rangle\}$ which are connected only through optical pumping, as shown in Fig. 5b. The master equation is then

$$\dot{\rho}_{ij} = -\frac{i}{\hbar}\left[\hat{H},\hat{\rho}\right]_{ij} + \frac{1}{2}\sum_{k=\{i,j\}}\sum_{l}\left[\delta_{ij}\gamma_{l\to k}\rho_{ll} - \gamma_{k\to l}\rho_{kk}\right], \tag{28}$$

where the Hamiltonian consists of $2\times 2$ blocks

$$\hat{H}_n = \begin{Bmatrix} \hbar\omega_2\left(n-\frac{1}{2}\right) + 2\hbar\gamma B_z & U_R\sqrt{n} \\ U_R^*\sqrt{n} & \hbar\omega_4\left(n+\frac{1}{2}\right) + 4\hbar\gamma B_z \end{Bmatrix}, \tag{29}$$

and the sum over $l$ includes only the non-zero relaxation rates indicated on Fig. 5b.

Optical pumping $|n, m=2\rangle \to |n', m=4\rangle$ is in principle provided by the lattice light field, but efficient cooling requires the addition of a separate $\sigma_+$ polarized pumper beam resonant with the $F=4 \to F'=4$ transition. This helps confine population to the $m=2,4$ manifolds. More importantly, the $m=4$ sublevel is dark with respect to this pumper beam, and we can achieve a pumping rate $\gamma_p >> \gamma_s$ without extra heating that would result from pumping on the $F=4 \leftrightarrow F'=5$ transition. The price of this arrangement is occasional optical pumping to the $F=3$ hyperfine manifold. Optical pumping to $F=3$ is problematic, because the optical potentials associated with the two hyperfine ground state manifolds are offset by $\lambda/4$, so an atom pumped into $F=3$ finds itself on the top of a potential hill. Unwanted heating can be avoided if the atom is repumped back into the $F=4$ manifold on a timescale short compared to the time in which wave packet disperses, of order $\sim 1/\omega_{osc}$. This is easily accomplished by adding also a $\sigma_+$ polarized repumper tuned to the $F=3 \to F'=4$ transition, which repumps atoms at a rate $\sim \Gamma >> \omega_{osc}$. Solving the rate equations for optical pumping we find that an average of ~2 pumper and ~1 repumper photons are scattered before the atom is returned to the stretched state. The three-step optical pumping process transfers an average mean-squared momentum along $\hat{z}$ of $21\hbar^2k^2/5$, which is three times the average momentum transfer in a single-step process. To first order in the small parameter $\eta^2$ and ignoring the small effect of curvature difference between the wells, the pumping rates from Fig. 5b then become

$$\gamma_{n\to n} \approx \gamma_p\left[1-\tfrac{21}{5}\eta^2 n\right], \qquad \gamma_{n\to n'\neq n} = \gamma_p \tfrac{21}{5}\eta^2 \max(n,n'). \tag{30}$$

For photons scattered from the lattice light field the mean-squared momentum transfer along $\hat{z}$ is $11\hbar^2k^2/15$, yielding

$$\gamma'_{n\to n} = \gamma_s\left(1-\tfrac{11}{15}\eta^2 n\right), \qquad \gamma'_{n\to n'\neq n} = \gamma_s \tfrac{11}{15}\eta^2 \max(n,n'), \tag{31}$$

for the remaining rates from Fig. 5b .

Sideband cooling can now be simulated by integrating the master equation, Eq. (28), with initial vibrational populations $\pi_n$, corresponding to a thermal state with Boltzman factor $q_B = \exp(-\hbar\omega_{osc}/k_B T) = \pi_{n+1}/\pi_n$. As an example, Fig. 6 shows the evolution of a system with single beam light shift $U_1 = 500\ E_R$, detuning $\Delta = -2000\,\Gamma$, initial Boltzman factor $q_B = 0.5$, and a sequence of cooling steps designed to transfer population $|n=5, m=4\rangle \rightarrow \ldots \rightarrow |n=0, m=4\rangle$. After these five steps substantial cooling has been achieved, and the ground state population has increased from the initial $\pi_0 = 0.5$ to $\pi_0 \approx 0.86$. Also, the task of producing a completely pure ground state by state selection has been much simplified, as high selectivity against the closest state $|n=1, m=4\rangle$ is no longer so critical.

The imperfect population transfer evident in Fig. 6 indicates below optimal Raman coupling strength, so that population escapes to highly excited states during the several cooling steps. The problem becomes more severe as we try to accumulate population from higher lying vibrational states; in the above example adding the last three steps to the cooling sequence resulted in a ground state population gain of only ~4%. Significant improvement in the cooling efficiency can be achieved with a modest increase of the figure-of-merit $\kappa'$, but this would require values $U_1 >> 500\ E_R$ which are probably not realistic. Limits on $U_1$ are imposed by the need to stay far detuned from the excited hyperfine manifold, and by the available laser power. Though a large figure of merit may be difficult to achieve in 1D lattices, it is readily available in 2D and 3D geometries. Sideband cooling in higher dimensional lattices will of course require a much more elaborate sequence of cooling steps, in part due to vibrational degeneracy, and in part due to the existence of non-coupled states within each of the degenerate vibrational manifolds [42]. A more detailed theoretical model of cooling in the 2D lattice geometry of Fig. 4 will be the subject of future work.

## III. Quantum Tunneling and Schrödinger Cats.

We have seen in Sec. II the flexibility with which one can prepare and coherently manipulate an atomic wave packet in an optical lattice. In this section we discuss how these techniques may be used to study atomic tunneling in an optical double-well potential, introduced in Eq. (3), when coupled to a noisy (but well characterized) environment. Though not macroscopic in the sense of a condensed matter system with on order Avagadro's number of particles, the separation between the wells of a given pair is on the order of the optical wavelength, which may be considered macroscopic when compared to the atom dimension. As such, an atom that is coherently distributed on two sides of the double wells may be considered to be a "Schrödinger cat".

To establish a clear physical picture, and elucidate the important scaling laws and order of magnitude of the effects, we first return to the simplified $|J=1/2\rangle \rightarrow |J'=3/2\rangle$ atom in a 1D lin-angle-lin lattice with an external transverse magnetic field applied along *x*. The resulting potential operator is given by Eq. (3). For a sufficiently deep potential, the vibrational energy level spacing is large compared to the Larmor frequency $\Omega_\perp = \gamma B_x$, in which case the magnetically induced coupling can be treated as a perturbation on the two harmonic wells for the two spin states $\{|m=\pm\rangle\}$,

$$\hat{U}(z) = \tfrac{1}{2} M\omega_{osc}^2\left[(z-\Delta z/2)^2|+\rangle\langle+| + (z+\Delta z/2)^2|-\rangle\langle-|\right] + \hbar\Omega_\perp\left(|+\rangle\langle-| \; + \; |+\rangle\langle-|\right), \quad (32)$$

where $\hbar\omega_{osc} = 2\sqrt{E_R U_p}$, and where $U_P$ and $\Delta z$ are given in Eq. (4). The ground state splitting arising from the coupling between the neighboring wells is approximately $\delta E \approx \hbar\Omega_\perp\langle 0_L | 0_R\rangle$, where $|0_L\rangle$ and $|0_R\rangle$ are the ground state wave function in the left and right potential wells having spin $|+\rangle$ and $|-\rangle$ respectively. In the harmonic approximation (i.e. Gaussian ground state wavefunctions),

$$\delta E \approx \hbar\Omega_\perp e^{-(k\Delta z)^2/8\eta^2} = \hbar\Omega_\perp e^{-\frac{1}{2}M\omega_{osc}^2(\Delta z/2)^2/\frac{1}{2}\hbar\omega_{osc}}. \quad (33)$$

The right-hand side shows that the ground state splitting scales exponentially with the ratio of the potential energy where the diabatic potentials cross (a measure of the barrier height), to the ground state energy in the unperturbed wells. For a typical experiment we might choose $\Delta z \approx \lambda/6$, in which case the separation between the double minima at one lattice site is half the distance to the neighboring site; in this case we can neglect tunneling between neighboring sites. Choosing $U_1 \approx 50\,\mathrm{E}_R$ this gives $\delta E \approx 0.1\,\hbar\Omega_L$ . For Cs the Bohr magneton is $\mu_B \approx 680\,E_R/\mathrm{G}$, and a moderate transverse magnetic field of 7 mG results in a ground state splitting of order one recoil energy.

An important practical consideration is inhomogeneous broadening of the tunneling resonance. Broadening results from changes in the optical well depth across the lattice volume, arising from variations in the laser intensity. Equation (33) implies a variation in ground state energy splitting

$$\frac{\Delta(\delta E)}{\delta E} \approx \frac{(k_L \Delta z)^2}{16\eta^2}\frac{\Delta U_1}{U_1}. \quad (34)$$

For the parameters above this implies $\Delta(\delta E)/\delta E \approx \Delta U_1/U_1$, i. e. even a relatively large variation of 10% in $U_1$ across the lattice volume will allow observation of ten coherent oscillations between the left and right potential wells.

The physical interpretation of these coherent oscillations is quite subtle, even for the seemingly simple potential of Eq. (32). A particle is said to *tunnel* between two wells if its total energy is less than the potential energy within the barrier separating them. In that case the motion between the wells is classically forbidden. Such a definition of tunneling is precise for a scalar particle, but for a nonseparable potential depending on the particle's internal degrees of freedom, the issue of classical versus non-classical motion is more complex, since the internal state may or may not adiabatically follow the center of mass motion. We will defer the formulation of a general definition of tunneling in this system to a future publication; here we adopt instead the following unambiguous, sufficient condition. The adiabatic potential

$$U_{adiab}(z) = \tfrac{1}{2} M\omega_{osc}^2\left(z^2 + \left(\tfrac{1}{2}\Delta z\right)^2\right) - \sqrt{\left(M\omega_{osc}^2 z\Delta z\right)^2 + \left(\hbar\Omega_\perp\right)^2}\ , \tag{35}$$

is the lowest eigenvalue of $\hat{U}(z)$ and therefore the lowest possible energy allowed at a given position. Energy conservation implies that the semiclassical motion is *always* bounded within one well (regardless of the initial conditions) if the total energy falls below the adiabatic barrier $U_{adiab}(0) = \frac{1}{8} M\omega_{osc}^2 \Delta z^2 - \hbar\Omega_\perp$, in which case we can interpret oscillations between the wells as a manifestation of quantum tunneling. For example, if we take $\hbar\Omega_\perp/E_R \approx 5 \Rightarrow \delta E/E_R \approx 0.5$ we find a barrier height $U_{adiab}(0) \approx 15.3$, while at the same time the energy of the states $|0_R\rangle$, $|0_L\rangle$ is $\hbar\omega_{osc}/2E_R \approx 8.6$, i. e. quantum tunneling through a classical barrier.

Though this two-level model serves to establish physical intuition, one must be cautious when scaling the expressions derived above to the real alkali atoms. Let us return to the case of Cs. In the very far off resonance limit, the atomic potential is given by Eq. (32) with the Pauli spin operators replaced by normalized hyperfine ground state angular momentum operators, $\hat{\sigma} \to \hat{\mathbf{F}}/F$ , in the expression for the light shift, and $\hat{\sigma}/2 \to \hat{\mathbf{F}}$ in the Zeeman interaction . In the presence of a transverse magnetic field we have a coupled set of nine diabatic potentials

$$\begin{aligned}\hat{U}(z) = \sum_m \left( -\frac{4}{3}U_1 - U_{p,m}\cos(2k_L z - k_L\Delta z_m) - \hbar\gamma m B_z \right) |F,m\rangle\langle F,m| \qquad (36)\\ -\frac{\hbar\gamma}{2}\left\{ \sqrt{F(F+1) - m(m+1)}(B_x - iB_y)|F,m+1/\langle F,m| + H.c \right\},\end{aligned}$$

with modulation depth , and nearest well separation

$$U_{P,m} = \frac{4}{3}U_1\sqrt{4\cos^2\theta + \left(\frac{m}{F}\right)^2 \sin^2\theta}\ , \qquad k_L\Delta z_m = \tan^{-1}\left(\frac{m\tan\theta}{2F}\right). \tag{37}$$

In the absence of a longitudinal field, making the harmonic approximation for the deeply bound states, the potential can be written as a coupled set of pairwise degenerate parabolic wells,

$$\hat{U}(z) \approx \sum_m U_{p,m} \Big\{ \big[ k_L^2 (z - \Delta z_m / 2)^2 - 1 \big] |F,m\rangle\langle F,m| + \big[ k_L^2 (z + \Delta z_m / 2)^2 - 1 \big] |F,-m\rangle\langle F,-m| \Big\} + V_{m\pm 1,m} |F, m \pm 1\rangle\langle F,m| \quad . \tag{38}$$

The transverse magnetic field will cause the atom to oscillate between the wells together with an oscillation of the atomic magnetization due to the Larmor precession. Note that the angular momentum will not generally oscillate over the full range between $|m=+F\rangle$ and $|m=-F\rangle$, since all magnetic sublevels are coupled, even at the bottom of the wells. The splitting of the ground state energy is not as simply approximated as it was for the spin 1/2 case, since the degenerate ground state doublet is coupled by no less than eighth order in the interaction potential. Perturbation theory quickly becomes intractable when the degeneracy is not broken in first or second order [43]. For this reason we go directly to a numerical solution for the fully quantum Hamiltonian, with potential Eq.(11).

The bandstructure, together with the three deepest adiabatic potentials, is shown in Fig. 7 with $U_1 = 150 E_R$, $\theta = \pi / 2.3$, and $\hbar\gamma B_T = 10\, E_R$. The ground band doublet is split by $\delta E \approx 1.8\, E_R$, and lies below the top of the adiabatic barrier; this satisfies the sufficient condition for quantum tunneling described above. For such tightly bound bands, there is negligible tunneling between different lattice sites, and each double well can be considered as isolated from its neighbors. In this case the Wannier spinor is well approximated by one period of the Bloch spinor,

$$\left| \psi_{n,q} \right\rangle = \sum_m e^{iqz} \left| u_{n,q}^{(m)} \right\rangle \otimes |F,m\rangle, \tag{39}$$

for any $q$. Fig. 7 shows the Bloch wave functions for the ground state doublet with $q = 0$, which are symmetric/antisymmetric according to $u_{n,q}^{(m)}(-z) = \pm u_{n,q}^{(-m)}(z)$. In the limit of large transverse magnetic fields, the barrier is removed and these states map onto the ground and excited states of the deepest adiabatic potential well. We define ground states localized in the left or right potential wells in terms of the symmetric and antisymmetric states $|S\rangle, |A\rangle$, in the usual way, $|L,R\rangle = \big(|S\rangle \pm |A\rangle\big) / \sqrt{2}$. The average magnetization of the localized states is calculated as

$$\left\langle \hat{F}_z \right\rangle_{L,R} = \sum_m m \int dz \left| u_{L,R}^{(m)}(z) \right|^2 . \tag{40}$$

For the parameters chosen above, this gives $\left\langle F_z \right\rangle_{L,R} = \pm 2.66$, which is sufficiently large to be resolved in real time.

Preparation of the states $|L,R,S,A\rangle$ can be accomplished using the techniques described in Sec. II. Through a combination of sideband cooling and state selection, the atoms are first prepared in a 1D lin⊥lin lattice in the "pure state" $|n=0, m=4\rangle$. A small longitudinal magnetic field $B_Z$ is applied to break the degeneracy between neighboring potential wells, thereby preventing the system from tunneling. The polarization is then adiabatically rotated to bring the wells together in pairs, and the transverse magnetic field is adiabatically ramped from zero to the value which produces the desired coupling. If $B_Z$ is then ramped to zero at a rate slow compared to the tunneling rate, the initial state will adiabatically connect to $|S\rangle$, as shown in Fig. 8. If on the other hand $B_Z$ is turned off rapidly compared to the tunneling rate, but slow compared to the energy spacing between the ground doublet and the next excited state, the atom will be prepared in $|R\rangle$. Each of these states will allow us to explore the effects of dissipation on quantum coherence.

To do so we must assure that decoherence caused by photon scattering is suppressed. Because coherent coupling in our system derives from the externally applied magnetic field (rather than the intrinsic Raman coupling discussed in Sec. II.B), this is easily accomplished by detuning very far from resonance. Note that photon scattering need not be fully suppressed, since its does not act to completely decohere a Schrödinger cat state, whose constituent $|L\rangle, |R\rangle$ wave packets are separated by less than $\lambda/2$ [44]. Once coherent tunneling has been accomplished, we can simulate the effects of coupling to an environment by introducing carefully designed temporal fluctuations of the lattice, with well defined statistical properties (noise spectrum, etc.). For example, fluctuations in the angle between the laser polarizations, $\theta \to \theta + \varepsilon(t)$, simulates phonons in the lattice, which adds a noise source to the Hamiltonian of the form

$$\hat{H}_{noise}(t) = \varepsilon(t)\frac{2U_1}{3}\left[2\sin\theta\cos(2k_L z)\hat{I} - \cos\theta\sin(2k_L z)\frac{\hat{F}_z}{F}\right]. \tag{41}$$

Fluctuations in the transverse magnetic field translate into fluctuations in the barrier height, which is an important component in the Bütteker-Landauer paradigm for exploring the time it takes a particle to traverse a forbidden barrier [45]. Time dependent lattice potentials can also be used to drive the double well system in a coherent fashion; the interplay between quantum tunneling and coherent drives then opens the door to exploration of a whole separate class of phenomena. For example, coherent oscillations of the longitudinal magnetic field can be used to drive the "bias", or energy asymmetry between the wells. In that case it has been predicted that certain frequencies will completely suppress tunneling due to quantum interference [46], a phenomenon known as “dynamic localization” [47].

## IV. Summary.

We have explored quantum state preparation and coherent control in a new physical system consisting of atoms bound in an optical lattice. In doing so our goal has been to exploit, as far as possible, the flexibility available in designing optical potentials. The potential operator in general has both diagonal elements, which are the diabatic potentials associated with different magnetic sublevels, and off-diagonal parts, which represent Raman coupling between magnetic sublevels. In a properly designed lattice we can trap atoms in nearly harmonic potential wells, and at the same time build in Raman coupling between the vibrational manifolds associated with pairs of magnetic sublevels. If the strength of this Raman coupling is modest compared to the vibrational spacing, then coupling between the manifolds can be controlled by applying a magnetic field parallel to the quantization axis, and shifting pairs of states into or out of degeneracy. When Raman coupling of appropriate strength and symmetry has been obtained, it is in principle straightforward to manipulate the atomic quantum state using the standard repertoire of $\theta$-pulses, adiabatic rapid passage etc., in analogy to the techniques demonstrated for trapped ions.

Special consideration has been given to the case of alkali atoms trapped in far-off-resonance optical lattices. In this situation it is useful to establish a basis independent representation of the optical potential. We have shown that in this limit one can separate the potential into a part independent of $F$ and $m$, and an effective magnetic field that can vary spatially on the scale of an optical wavelength. Our analysis shows that Raman coherences of the form $|F,m\rangle \leftrightarrow |F,m\pm 1\rangle$ persist in the infinite detuning limit, provided that the lattice light field can be designed to contain both σ and π-polarized components. The flexibility to do so is available in 2D and 3D lattice configurations. We define a figure-of-merit for state manipulation, $\kappa \equiv U_R/\hbar\gamma_s$, which can fall in the range 10-100. At large but finite detuning Raman coherences of the type $|F,m\rangle \leftrightarrow |F,m\pm 2\rangle$ also occur, but vanish asymptotically as $U_R \propto \eta\, U_1/\Delta$. These coherences are therefore driven at a rate comparable to the rate of decay due to spontaneous light scattering, and are not useful for state manipulation in the alkalis.

Preparation of a pure quantum state can be achieved through resolved-sideband Raman cooling based on coupling terms intrinsic to the lattice. The requirements for resolved sideband cooling, $\hbar\omega_{osc} >> U_R >> \hbar\gamma_s(\Delta k z_0)^2$, are much less restrictive than for coherent state manipulation. A simple model shows that sideband cooling via the $|F,m\rangle \leftrightarrow |F,m\pm 2\rangle$ coupling in a 1D lin⊥lin lattice can bring the system quite close to a pure state. In higher dimensional lattices one can use $|F,m\rangle \leftrightarrow |F,m\pm 1\rangle$ type Raman coupling. Defining a figure-of-merit for cooling, $\kappa' \equiv U_R \big/ \hbar\gamma_s(\Delta k z_0)^2$, we find that it can easily be of order $10^3$ in a representative 2D configuration.

Quantum tunneling in a double-well potential is an important paradigm for coherent evolution in quantum systems coupled to a noisy environment. A 1D lin-angle-lin far-off-resonance lattice with transverse magnetic field provides an array of double-well potentials that can be used for controlled experiments of this type. We have performed a band-theoretical analysis for Cs atoms in this potential. Our results indicate that it is possible to find parameter regimes where several coherent tunneling oscillations should be observable in real-time and where the tunneling coupling can be used to prepare both localized ($|L,R\rangle$) and Schrödinger cat states ($(|L\rangle \pm |R\rangle)\big/\sqrt{2}$. Once coherent evolution and control has been achieved, detailed studies of decoherence can be carried out through the introduction of well characterized fluctuations in the lattice potentials.

We acknowledge helpful discussions with D. J. Heinzen, C. M. Caves, and P. M. Alsing. This work was supported by NSF Grant No. Phy-9503259, by ARO Grant No. DAAG559710165, and by JSOP Grant No. DAAG559710116.

**Appendix A.**

In this appendix we derive Eq. (6) for the tensor polarizability of an alkali atom, which has been optically pumped into a given hyperfine ground state connected to a multiplet of hyperfine excited states. According to Eq. (5), in the spherical basis,

$$\hat{\alpha}_{q',q} = -\sum_{F',m} \frac{|F,m+q-q'\rangle\langle F,m+q-q'|\hat{d}_{-q'}|F',m+q\rangle\langle F',m+q|\hat{d}_q|F,m\rangle\langle F,m|}{\hbar\Delta_{F,F'}}, \tag{A1}$$

where we have invoked the dipole selection rules to reduce the sums in the projectors. According to the Wigner-Eckart theorem

$$\langle F',m+q|\hat{d}_q|F,m\rangle = c_{F,m}^{F',m+q} \left\langle F'(J',I)\middle\| d\otimes\hat{1}_I \middle\| F(J,I)\right\rangle, \tag{A2}$$

where $c_{F,m}^{F',m+q}$ is the Clebsch-Gordan coefficient for the dipole transition $|F,m\rangle \to |F',m+q\rangle$ , we have use the Condon and Shortley normalization for the reduced matrix element, and $\hat{1}_I$ is the unit operator on the nuclear spin subspace. The coupling law for the reduced matrix element gives,

$$\left|\left\langle F'(J',I)\middle\| d\otimes\hat{1}_I \middle\| F(J,I)\right\rangle\right|^2 = f_{F'F}\left|\langle J'\|d\|J\rangle\right|^2, \tag{A3a}$$

where

$$f_{F'F} = (2J'+1)(2F+1)\left|\begin{Bmatrix} F' & I & J' \\ J & 1 & F \end{Bmatrix}\right|^2 \tag{A3b}$$

is the relative oscillator strength for decay $|F\rangle \to |F'\rangle$, satisfying the sum rule $\sum_F f_{F',F} = 1$. Substituting Eqs. (A2-3) into (A1),

$$\hat{\alpha}_{q',q} = \tilde{\alpha} \sum_{F'} \frac{\Delta_{F_{\max},F'_{\max}}}{\Delta_{F,F'}} f_{F'F} \sum_m c_{F,m+q-q'}^{F',m+q}\, c_{F,m}^{F',m+q}\; |F,m+q-q'\rangle\langle F,m|, \tag{A4}$$

where

$$\tilde{\alpha} = \frac{\left|\left\langle J' \| d \| J \right\rangle\right|^2}{\hbar \Delta_{F_{\max}, F'_{\max}}} \tag{A5}$$

is the characteristic polarizability of the $|J\rangle \rightarrow |J'\rangle$, transition.

## Appendix B.

Consider the Hermitian part of the linear tensor polarizabilty operator, Eq. (5), for a $|J = 1/2\rangle \rightarrow |J' = 3/2\rangle$ transition. Normalized with respect to the characteristic polarizability scalar,

$$\overline{\alpha}_{ij} \equiv \frac{\hat{\alpha}_{ij}}{\tilde{\alpha}} = \hat{D}_i^\dagger \hat{D}_j, \tag{B1}$$

where we have defined normalized "creation" and "annihilation" dipole operators,

$$\hat{D}_i \equiv \frac{P_{J'} \hat{d}_i P_J}{\left\langle J' \| d \| J \right\rangle} = \sum_{m,q} \vec{\mathbf{e}}_i \cdot \vec{\varepsilon}_q^{\,*} \, c_m^{m+q} \, |J', m+q\rangle\langle J, m|. \tag{B2}$$

Since $\overline{\alpha}_{ij}$ acts on a two dimensional Hilbert space, decomposition into irreducible tensor operators must truncate at the vector term

$$\overline{\alpha}_{ij} = \frac{1}{3}\delta_{ij} \, \mathrm{Tr}\left(\hat{\mathbf{D}}^\dagger \cdot \hat{\mathbf{D}}\right) \frac{\hat{I}}{2} + \frac{1}{2}\varepsilon_{ijk} \left(\hat{\mathbf{D}}^\dagger \times \hat{\mathbf{D}}\right)_k. \tag{B3}$$

By symmetry,

$$\mathrm{Tr}\left(\hat{\mathbf{D}}^\dagger \cdot \hat{\mathbf{D}}\right) = 3\,\mathrm{Tr}\left(\hat{D}_z^\dagger \hat{D}_z\right) = 3\left(\langle + |\hat{D}_z^\dagger \hat{D}_z| + \rangle + \langle - |\hat{D}_z^\dagger \hat{D}_z| - \rangle\right) = 4. \tag{B4}$$

The vector part is anti-Hermitian, and thus must be of the form

$$\left(\hat{\mathbf{D}}^\dagger \times \hat{\mathbf{D}}\right)_k = iC\,\hat{\sigma}_k \;, \tag{B5}$$

where $C$ is some real constant, independent of $k$, which can be found by direct expansion,

$$\left(\hat{\mathbf{D}}^\dagger \times \hat{\mathbf{D}}\right)_z = -i\left(\hat{D}_+^\dagger \hat{D}_+ - \hat{D}_-^\dagger \hat{D}_-\right) = -\frac{2}{3} i\hat{\sigma}_z . \tag{B6}$$

Thus, the polarizability tensor for the $|J = 1/2\rangle \rightarrow |J' = 3/2\rangle$ has the form given in Eq. (9),

$$\hat{\alpha}_{ij}(J \rightarrow J') = \tilde{\alpha}\left(\frac{2}{3}\delta_{ij}\,\hat{I} - \frac{i}{3}\varepsilon_{ijk}\,\hat{\sigma}_k\right) \tag{B7}$$

## References

[1] *Mode Selective Chemistry*, ed. J. Jortner, R. D. Levine, and B. Pullamn (Dordrecht, Boston, 1991); M. Shapiro and P. Brunner, Annu. Rev. Phys. Chem. **43**, 257 (1992); W. Warren, H. Rabitz, and M. Dahleh, Science **259**, 1581 (1993).

[2] H. J. Kimble in *Fundamental Sytems in Quantum Optics. Les Houches Summer School, Session LIII*, eds. J. Dalibard, J.M. Raymond, and J. Zinn-Justin, p. 545 (Elsevier, Amsterdam, 1992); S. Haroche *ibid.* p. 771; H. Walther, Adv. At. Mol. Opt. Phys. **32**, 379 ( 1994).

[3] D. T. Smithey, M. Beck, M. G. Raymer and A. Faridani, Phys. Rev. Lett. **70**, 1244 (1993); U. Leonhardt and H. Paul, Prog. Quant. Elect. **19**, 89 (1995) and references therein.

[4] K. Vogel, V. M. Akulin and W. P. Sleich, Phys. Rev. Lett. **71**, 1816 (1993); A. Parkins, P. Marte, P. Zoller and H. J. Kimble, *ibid*, 3095 (1993); M. Brune, E. Hagley, J. Dreyer, X. Maitre, A. Maali, C. Hunderlich, J. M. Raimond and S. Haroche, *ibid* **77**, 4887 (1996).

[5] Z. D. Gaeta, M. W. Noel and C. R. Stroud, Phys. Rev. Lett. **73**, 636 (1994); M. W. Noel and C. R. Stroud, *ibid* **77**, 1913 (1996).

[6] F. Diedrich, J. C. Bergquist, W. M. Itano and D. J. Wineland, Phys. Rev. Lett. **62**, 403 (1989); C. Monroe, D. M. Meekhof, B. E. King, S. R. Jefferts, W. M. Itano, D. J. Wineland and P. Gould, *ibid* **75**, 4011 (1995).

[7] D. M. Meekhof, C. Monroe, B. E. King, W. M. Itano and D. J. Wineland, Phys. Rev. Lett. **76**, 1796 (1996); C. Monroe, D. M. Meekhof, B. E. King and D. J. Wineland, Science **272**, 1131 (1996).

[8] J. I. Cirac and P. Zoller, Phys. Rev. Lett. **74**, 4091 (1995); C. Monroe, D. M. Meekhof, B. E. King, W. M. Itano and D. J. Wineland, *ibid* **75**, 4714 (1995).

[9] P. Domokos, J.M. Raimond, M. Brune and S. Haroche, Phys. Rev. A **52**, 3554 (1995); Q. A. Turchette, C. J. Hood, W. Lange, H. Mabuchi, H. J. Kimble, Phys. Rev. Lett. **75**, 4710 (1995).

[10] D. P. DiVincenzo, Science **270**, 255 (1995); A. Ekert and R. Jozsa, Rev. Mod. Phys. **68,** 733 (1996) and references therein.

[11] A. O. Caldiera and A. J. Leggett, Phys. Rev. A **31**, 1059 (1985). W. H. Zurek, Phys. Today **44**(10), 36 (1991).
[12] See for example A. J. Legget in *Quantum Tunnelling in Condensed Media*, chap. 1, eds. Yu. Kagan and A. J. Leggett, (Elsevier Science Publishers B. V., Amsterdam, 1992).
[13] For a review of optical lattices, see P. S. Jessen and I. H. Deutsch, Adv. At. Mol. Opt. Phys. **37**, 95 (1996); G. Grynberg and C. Triché, in *Coherent and Collective Interactions of Particles and Radiation Beams*, ed. by A. Aspect, W. Barletta, and R. Bonifacio, International School of Physics "Enrico Fermi", CXXXI Course (ETS Editrice, Pisa, to be published) and references therein.
[14] P. Verkerk, B. Lounis, C. Salomon and C. Cohen-Tannoudji, Phys. Rev. Lett. **68**, 3861 (1992); P. S. Jessen, C. Gertz, P. D. Lett, W. D. Phillips, S. L. Rolston, R. J. C. Spreeuw and C. I. Westbrook, Phys. Rev. Lett. **69**, 49 (1992), A. Hemmerich and T. W. Hänsch, Phys. Rev. Lett. **70**, 410 (1993).
[15] D. L. Haycock, S. E. Hamann, G. Klose and P. S. Jessen, Phys. Rev. A **55**, R3991 (1997).
[16] F. L. Moore, J. C. Robinson, C. Bharucha, P. E. Williams and M. G. Raizen, Phys. Rev. Lett. **73**, 2974 (1994).
[17] M. B. Dahan, E. Peik, J. Reichel, Y. Castin and C. Salomon, Phys. Rev. Lett **76**, 4508 (1996); S. R. Wilkinson, C. F. Bharucha, K. W. Madison, Qian Niu and M. G. Raizen, *ibid* **76**, 4512 (1996).
[18] D. J. Heinzen and D. J. Wineland, Phys. Rev. A **42**, 2977 (1990).
[19] D. J. Wineland, J. J. Bollinger, W. M. Itano and D. J. Heinzen, Phys. Rev. A **50**, 67 (1992), and references therein.
[20] C. F. Bharucha, K. W. Madison, P. R. Morrow, S. R. Wilkinson, B. Sundaram and M. G. Raizen, Phys. Rev. A **55**, R857 (1997).
[21] S. R. Wilkinson, C. F. Bharucha, M. C. Fischer, K. W. Madison, P. R. Morrow, Q. Niu, B. Sundaram and M. G. Raizen, Nature **387**, 575 (1997).
[22] A. J. Legget, S. Chakravarty, A. T. Dorsey, M. P. A. Fisher, A. Garg and W. Zwerger, Rev. Mod. Phys. **59**, 1 (1987) and references therein; U. Weiss, "Quantum Dissipative Systems", (World Scientific, Singapore, 1993).
[23] A. O. Caldiera and A. J. Leggett, Phys. Rev. Lett. **46**, 211 (1981).
[24] J. Clarke, A. N. Cleland, M. H. Devoret, D. Esteve and J. M. Martinis, Science **239**, 953 (1988).
[25] P. Hänggi, P. Talkner and M. Borkovec, Rev. Mod. Phys. **62**, 251, (1990).
[26] A. Würger, "From Coherent Tunneling to Relaxation", Springer (1997).
[27] *Atom Interferometry* , ed. P. R. Berman (Academic Press, San Diego, 1997).
[28] J. Lawall, S. Kulin, B. Saubamea, N. Bigelow, M. Leduc and C. Cohen-Tannoudji, Phys. Rev. Lett. **75**, 4194 (1995).

[29] J. I. Cirac, A. S. Parkins, R. Blatt and P. Zoller, Adv. At. Mol. Opt. Phys. **37**, 95 (1996), and references therein.
[30] M. Kasevich and S. Chu, Phys. Rev. Lett. **69**, 1741 (1992); J. Lawall and M. Prentiss, Phys. Rev. Lett. **72**, 993 (1994); L. S. Goldner, C. Gerz, R. J. C. Spreeuw, S. L. Rolston, C. I. Westbrook, W. D. Phillips, P. Marte and P. Zoller, *ibid* **72**, 997 (1994).
[31] R. Taïeb, R. Dum, J. I. Cirac, P. Marte and P. Zoller, Phys. Rev. A **49**, 4876 (1994).
[32] V. Finkelstein, P. R. Berman, and J.Guo, Phys. Rev. A **45**, 1829 (1992).
[33] R. Taieb, P. Marte, R. Dum and P. Zoller, Phys. Rev. A **47**, 4986 (1993).
[34] W. Happer and B. S. Mathur, Phys. Rev. **163**, 12 (1967).
[35] Y. Castin and J. Dalibard, Europhys. Lett. **14**, 761 (1991); P. Marte, R. Dum, R. Taieb and P. Zoller, Phys. Rev. Lett **47**, 1378 (1993); M. Doery, M. Widmer, J. Bellanca, E. Vredenbregt, T. Bergeman and H. Metcalf, Phys. Rev. Lett. **72**, 2546 (1994).
[36] I. H. Deutsch, J. Grondalski, and P. M. Alsing, Phys. Rev. A **56**, R1705 (1997).
[37] J.-Y. Courtois, Annales de Physique **21**, 1 (1996).
[38] G. Grynberg, B. Lounis, P. Verkerk, J.-Y. Courtois and C. Salomon, Phys. Rev. Lett. **70**, 2249 (1993).
[39] B. P. Anderson, T. L. Gustavson and M. A. Kasevich, Phys. Rev. A 53, R3727 (1996).
[40] T. Müller-Seydlitz, M. Hartl, B. Brezger, H. Hansel, C. Keller, A. Schnetz, R. J. C. Spreeuw, T. Pfau and J. Mlyneck, Phys. Rev. Lett. **78**, 1038 (1997).
[41] D. L. Haycock, S. E. Hamann, G. Klose, G. Raithel and P. S. Jessen, to appear in Phys. Rev. A **57**, number 2 (February 1998).
[42] D. S. Weiss S. L. Winoto and M. T. DePue, Proc. SPIE **2995**, 156 (1997).
[43] See for example K. Gottfried, *Quantum Mechanics*, p. 397 (W.A. Benjamin, New York, 1966).
[44] M. S. Chapman, T. D. Hammond, A. Lenef, J. Schmiedmayer, R. A. Rubenstein, E. Smith and D. E. Pritchard, Phys. Rev. Lett. **75**, 3783 (1995); J. I. Cirac, A. Schenzle and P. Zoller, Europhys. Lett. **27**, 123 (1994).
[45] M. Büttiker and R. Landauer, Phys. Rev. Lett. **49**, 1739 (1992).
[46] T. Dittrich, F. Grossmann, P. Jung, B. Oelschlagel and P. Hanggi, Physica A **194**, 173 (1993).
[47] D. H. Dunlap and V. M. Kenkre, Phys. Rev. B **34**, 3625 (1986).

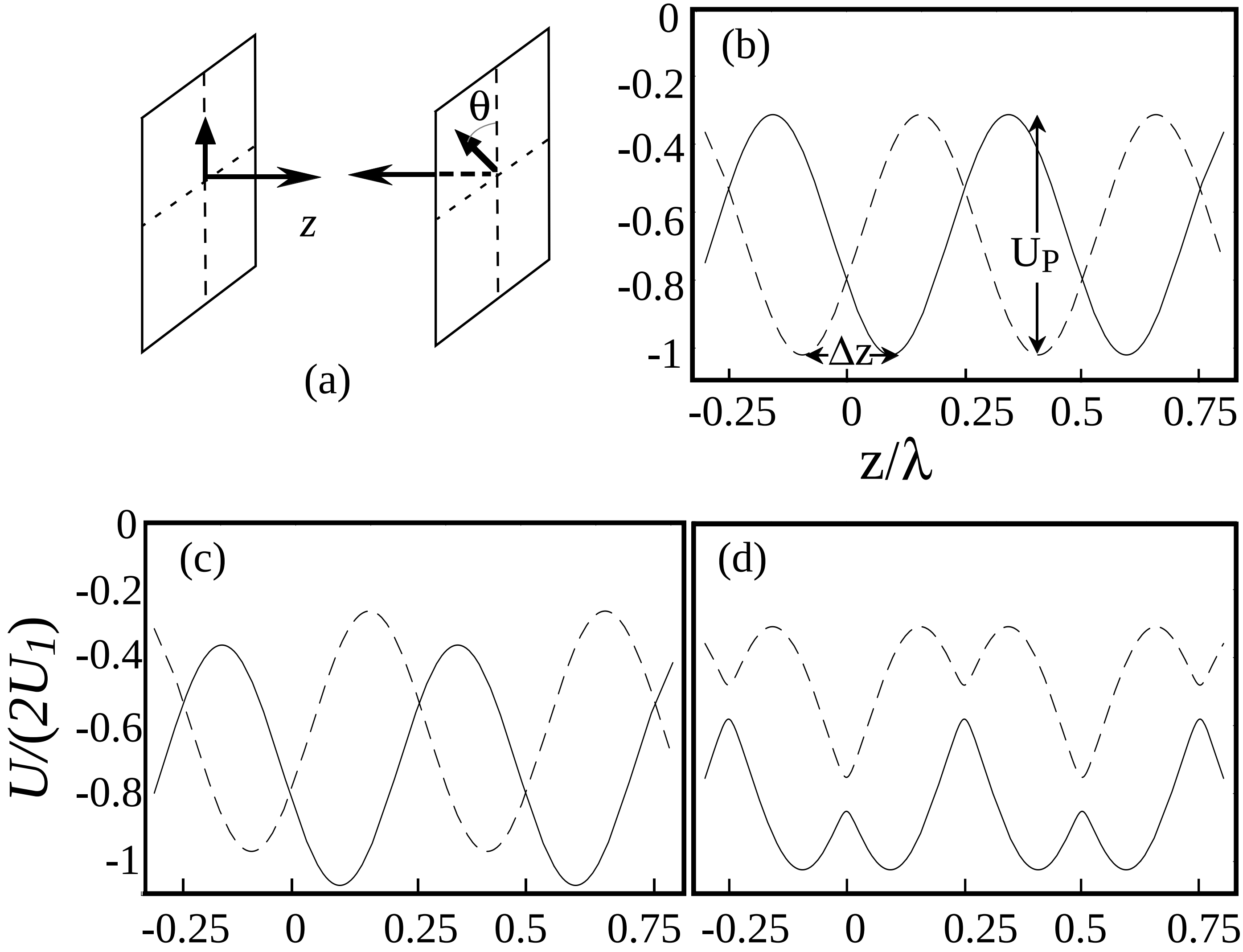

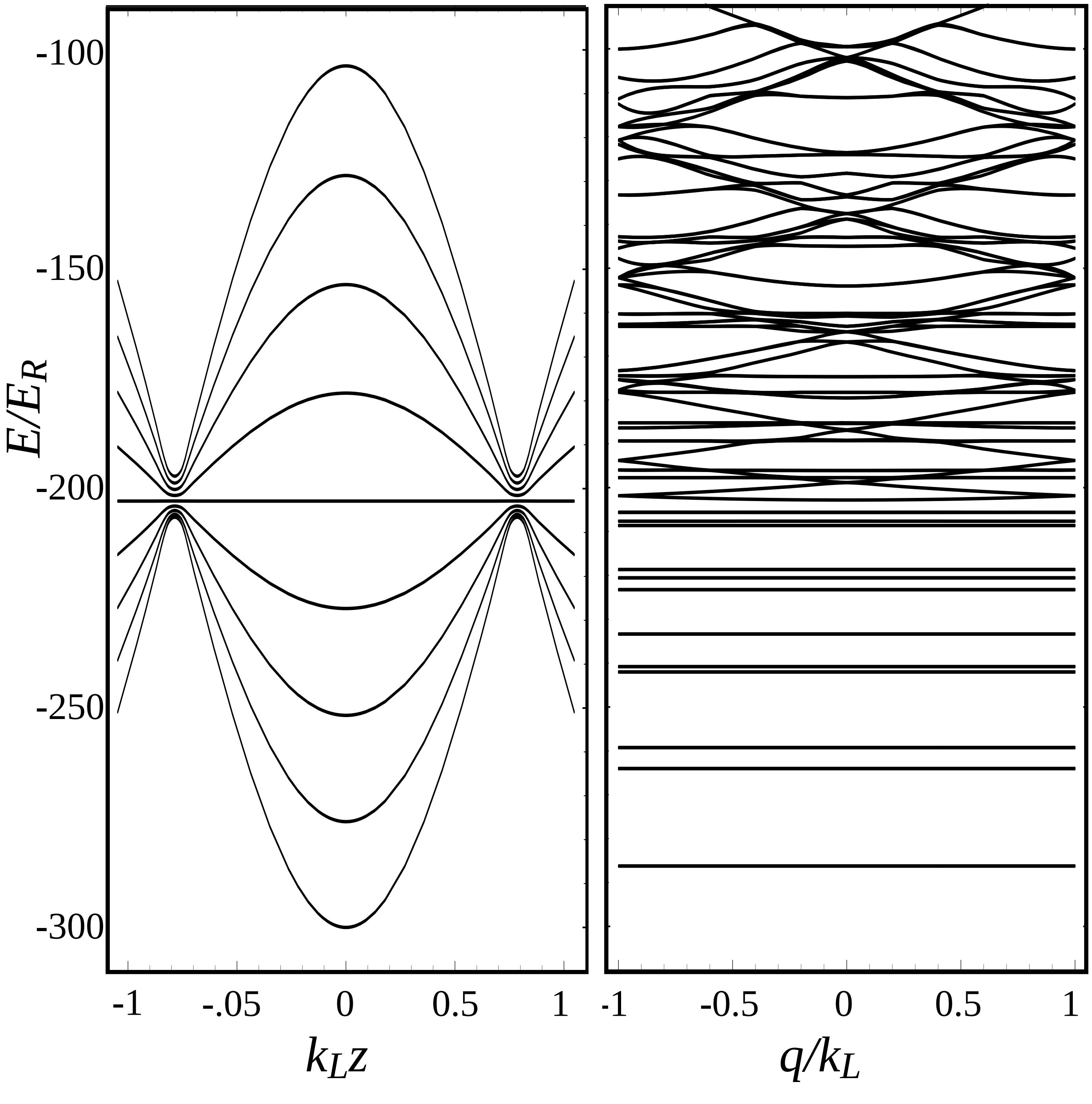

(a) (b)

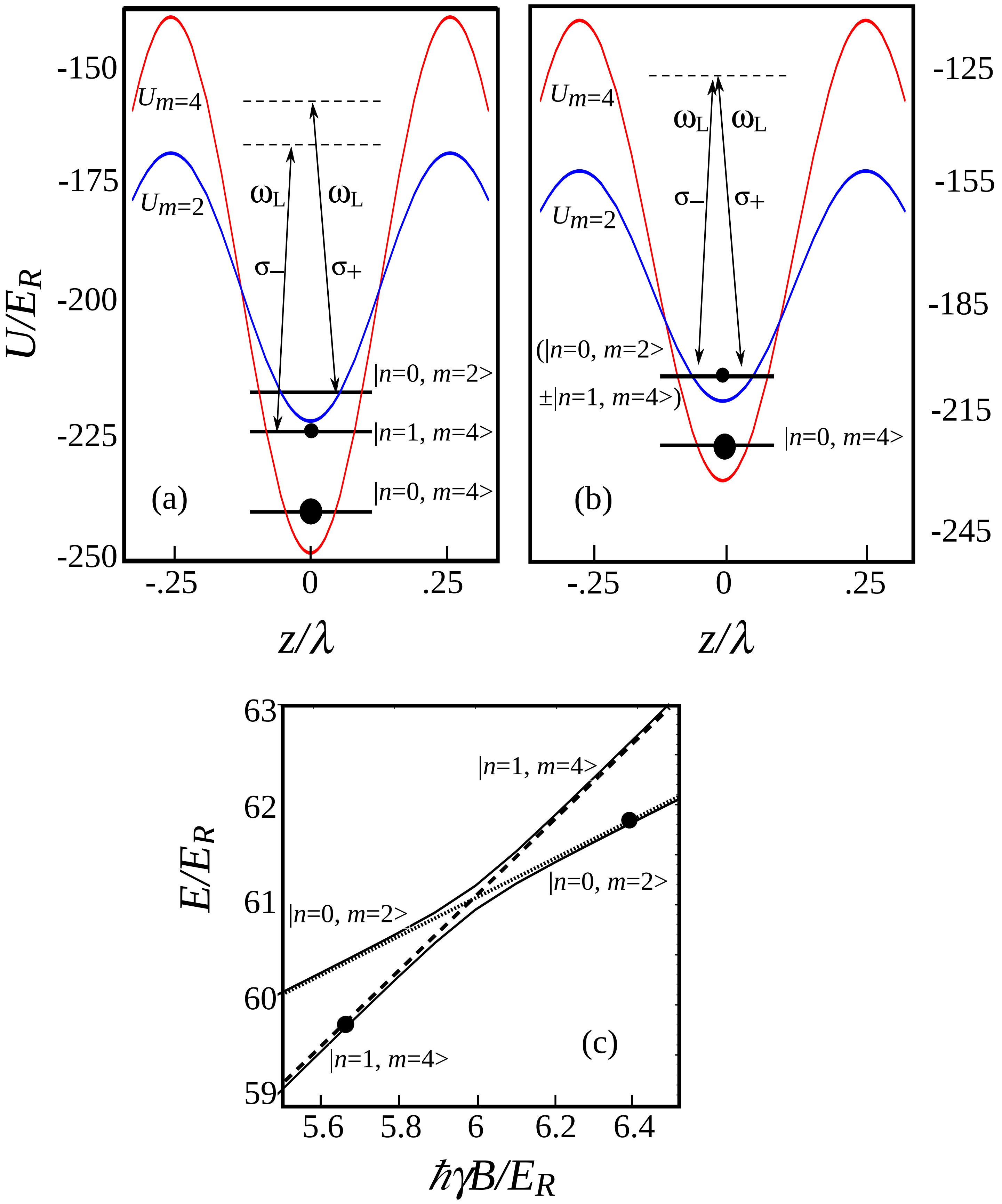

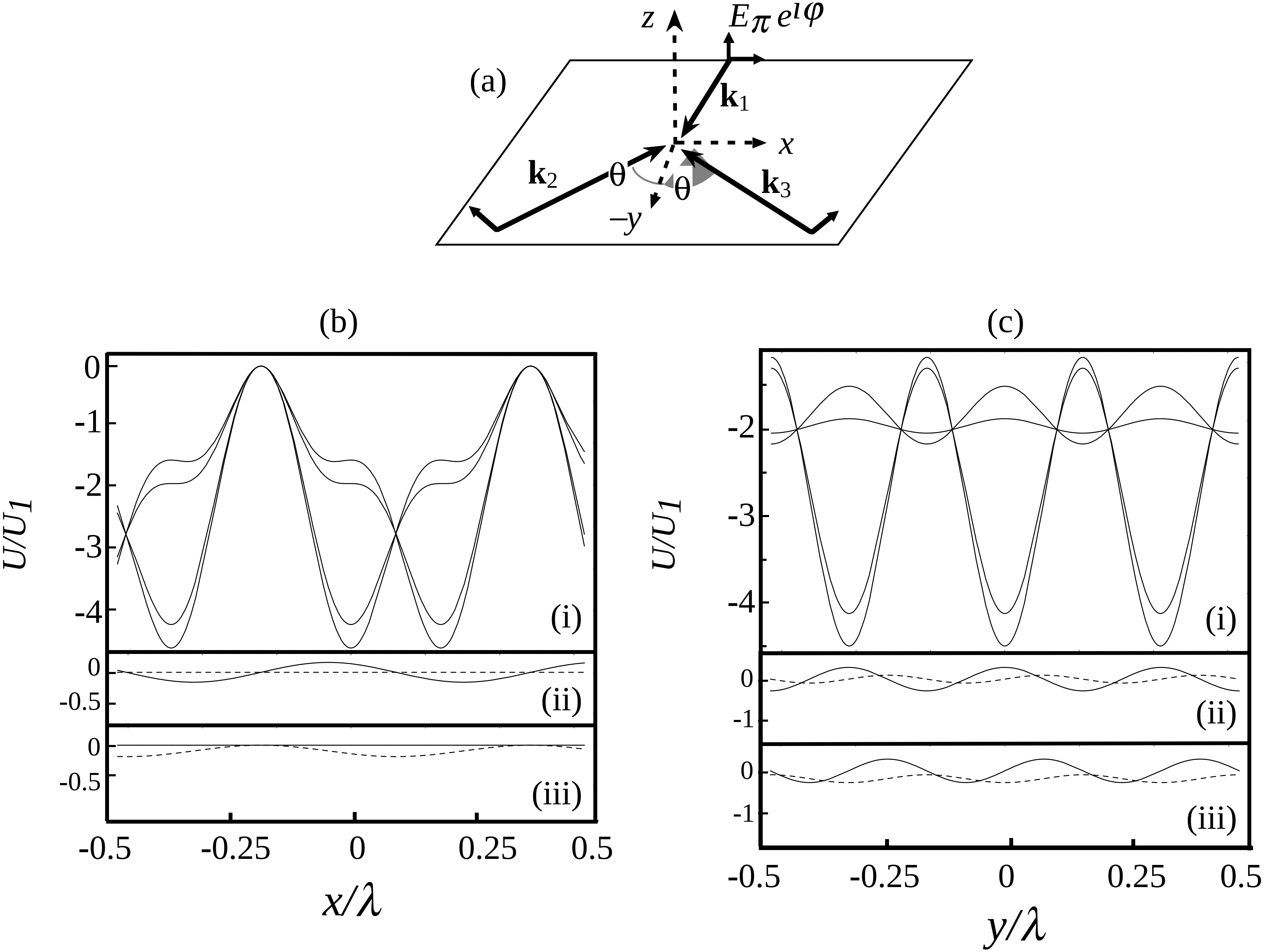

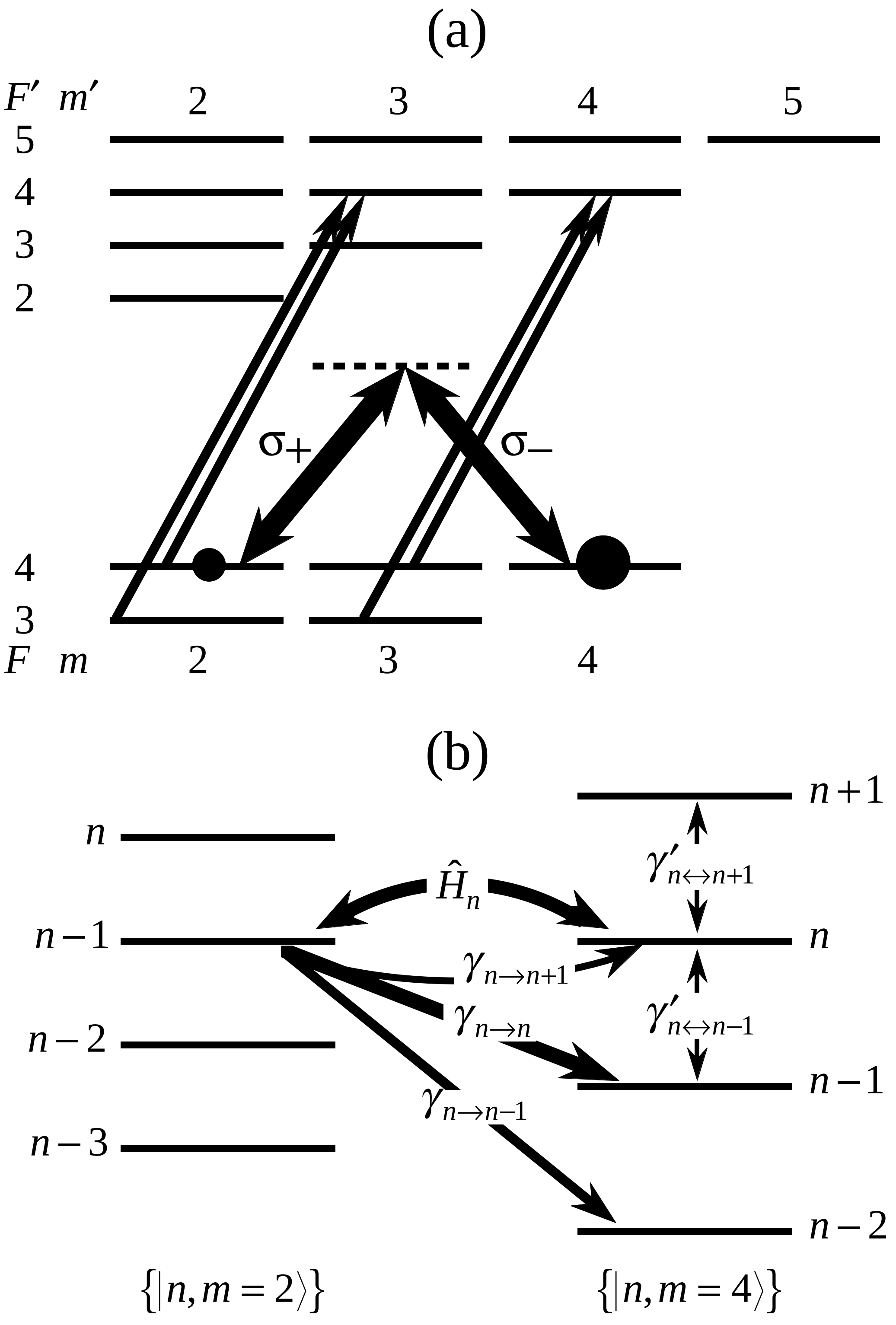

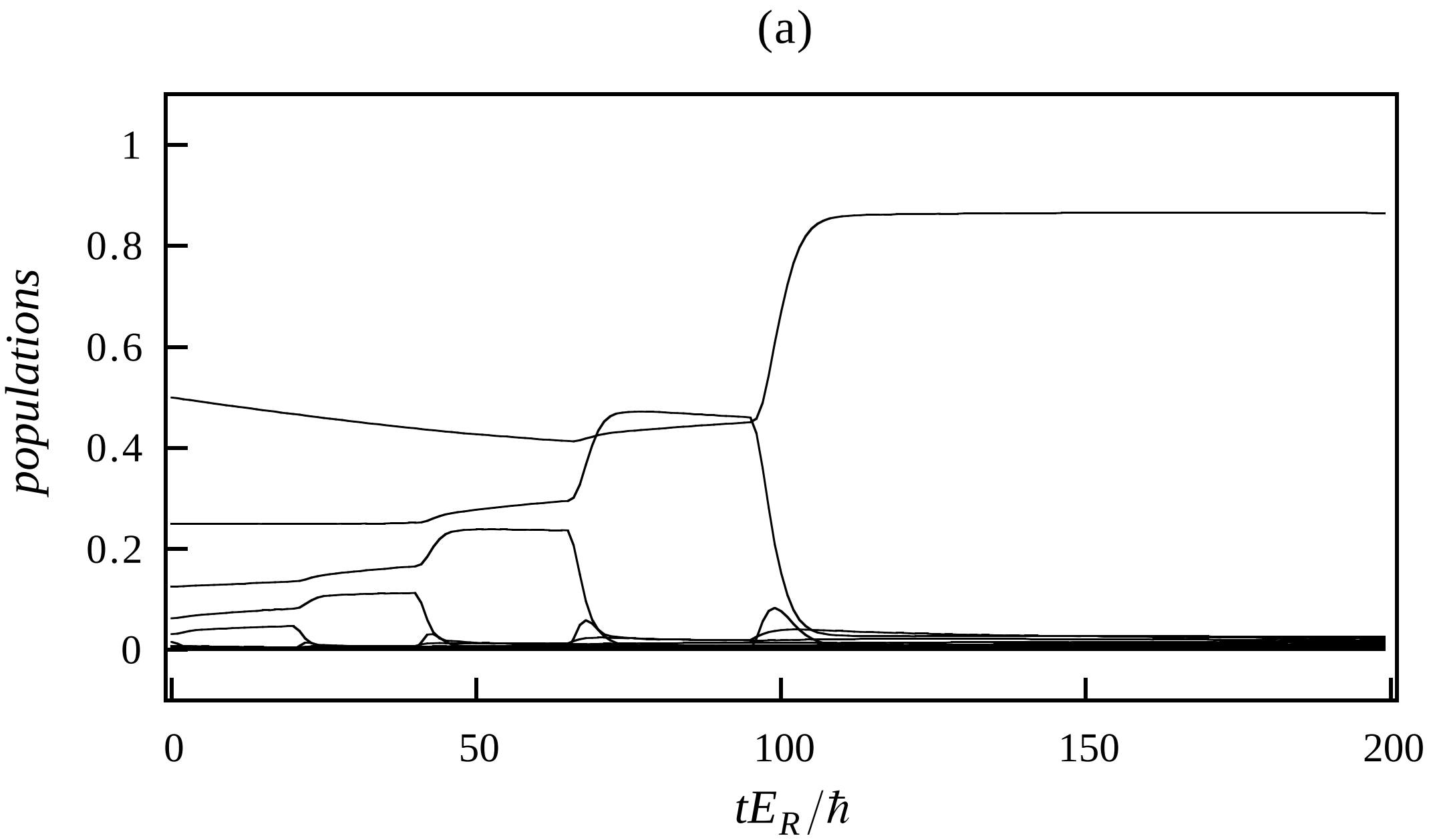

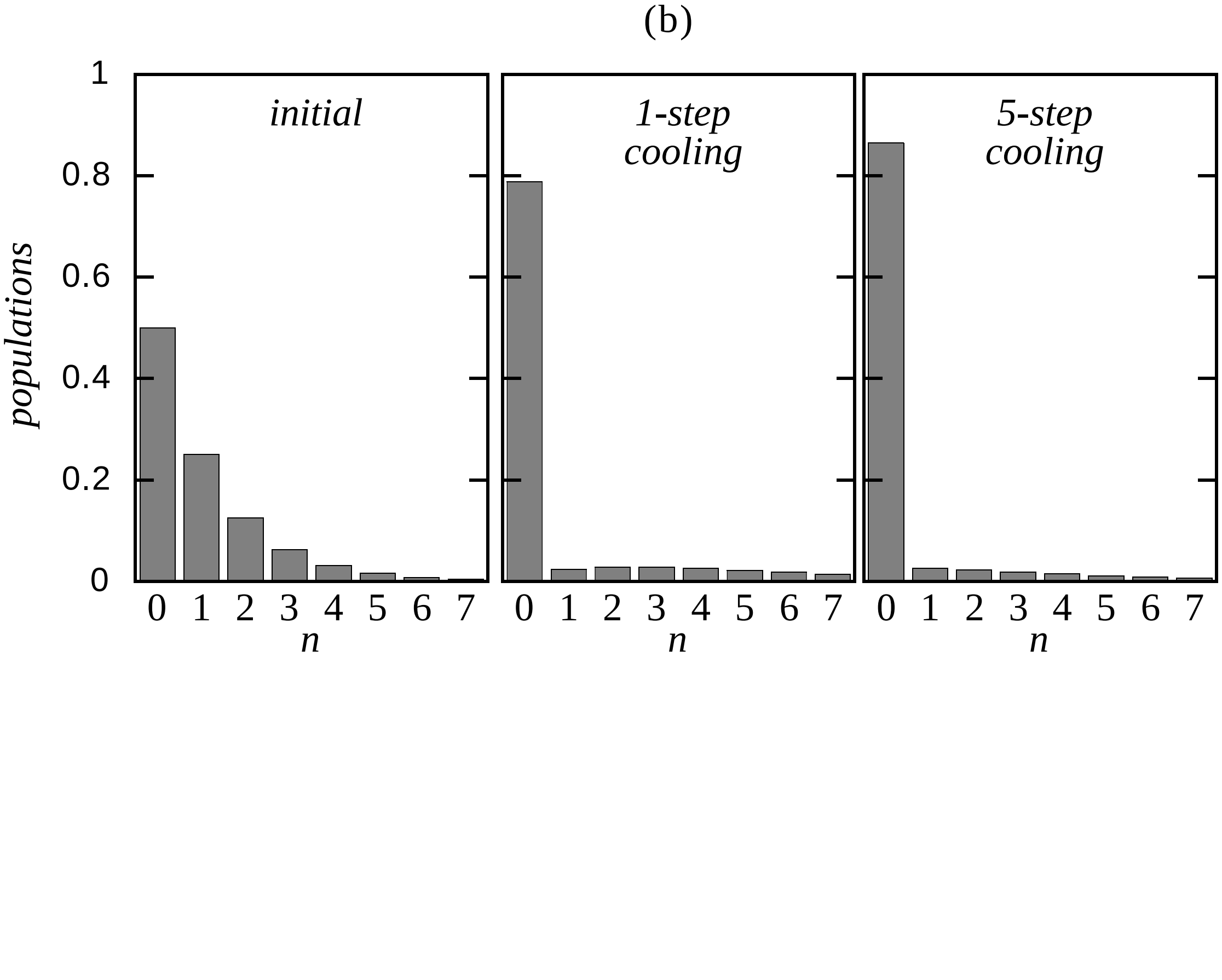

(b)
1
0.8
0.6
0.4
0.2
0
populations
initial
1-step cooling
5-step cooling
0 1 2 3 4 5 6 7
n
0 1 2 3 4 5 6 7
n
0 1 2 3 4 5 6 7
n

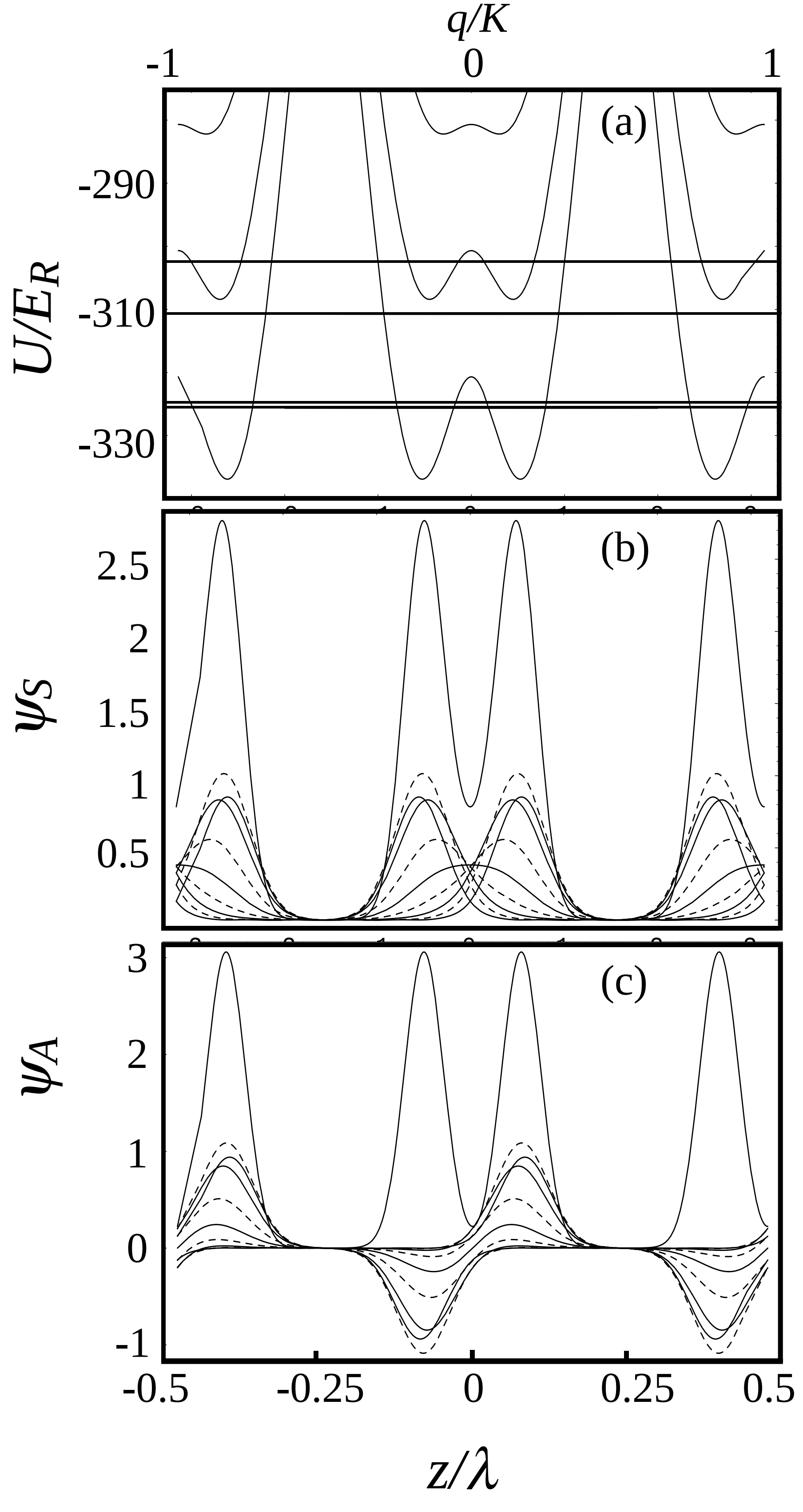

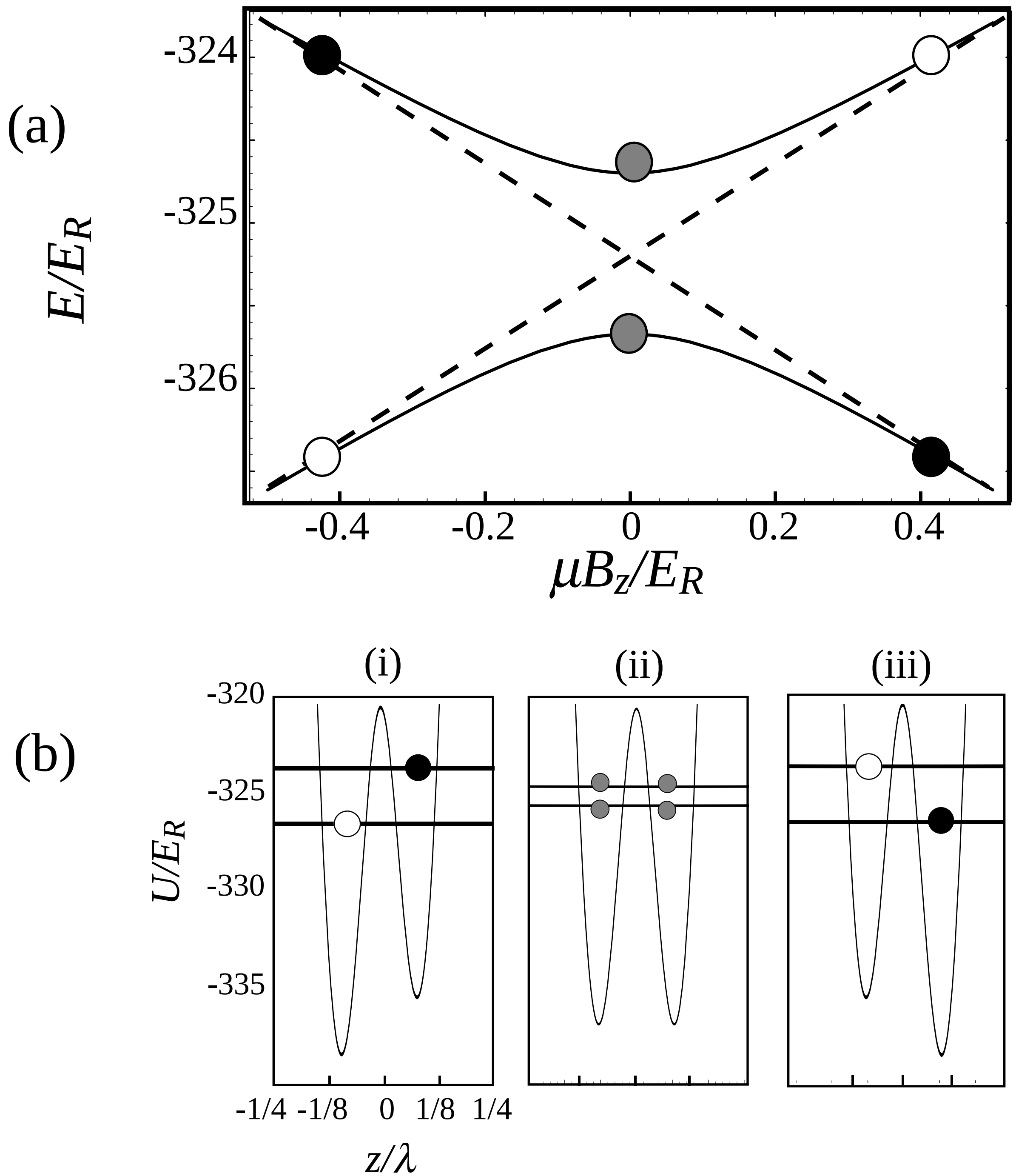